\documentclass[pre,twocolumn,superscriptaddress,longbibliography,floatfix]{revtex4-2}
\usepackage{graphicx}
\usepackage{amsmath,amssymb}
\usepackage{bm}
\usepackage{url}
\usepackage[colorlinks=true,linkcolor=blue,citecolor=blue,urlcolor=blue]{hyperref}

\newcommand{\pc}{p_c}
\newcommand{\ps}{p_c^{\rm site}}
\newcommand{\pb}{p_c^{\rm bond}}
\newcommand{\zb}{\bar z}

\begin{document}
\raggedbottom

\title{Site and bond percolation thresholds of all $k$-uniform tilings with $k\le 10$}

\author{Maksim Elisov}
\email{maksimelisov2003@gmail.com}
\homepage{ORCID: 0009-0001-3097-2703}
\affiliation{Samara National Research University, 34 Moskovskoye shosse, Samara 443086, Russia}

\date{\today}

\begin{abstract}
We determine the site and bond percolation thresholds of all 23\,396 $k$-uniform tilings of the plane by regular polygons with $k\le 10$, the complete set according to the known enumeration. We use Monte Carlo simulation on tori and the condition that no wrapping and a cross-wrapping cluster are equally probable, the Monte Carlo form of the critical-polynomial criterion, which needs no knowledge of the large-scale anisotropy that most of these lattices have. The 486 thresholds with $k\le4$ have uncertainties of a few units in the sixth decimal and reproduce the 22 Archimedean reference values (rms deviation 0.71 standard errors) and unpublished bond thresholds of the 20 two-uniform lattices ($\chi^2=19.5$ for 20 values); the others have median statistical uncertainties of $2\times10^{-5}$ ($k\le7$) to $7\times10^{-5}$ ($k\ge8$) and no detectable finite-size bias. The vertex-type fractions alone fix the thresholds for $k\ge5$ to about $2\times10^{-3}$ (linear rule fitted on $k\le4$), while tilings with equal composition differ by up to $6\times10^{-3}$ (site) and $2\times10^{-2}$ (bond). In the range of domain sizes covered, descriptors of radius one account for more than half of this difference, and stackings of identical strips have thresholds linear in the fraction of each boundary type. For tilings that differ only by the registry of strips separated by $h$ rows of squares, the bond threshold difference decreases by a factor of 6--8 per row, comparable to the leading momentum-$\pi$ eigenvalue of the square-band transfer matrix.
\end{abstract}

\maketitle

\section{Introduction}

The percolation threshold $\pc$ is the simplest lattice-dependent quantity of percolation theory, and the one most often needed in applications~\cite{StaufferAharony,Grimmett99}. In two dimensions exact values are known for a few lattices: $1/2$ for site percolation on the triangular lattice~\cite{Kesten82} and for bond percolation on the square lattice~\cite{Kesten80}, $2\sin(\pi/18)$ and $1-2\sin(\pi/18)$ for bond percolation on the triangular and honeycomb lattices~\cite{SykesEssam64,Wierman81}, and further values that follow from star--triangle and self-duality arguments~\cite{SykesEssam64,ZiffScullard06,ScullardZiff08,WiermanZiff11,GrimmettManolescu14}. For all other lattices $\pc$ is known numerically, from Monte Carlo simulation~\cite{NewmanZiff00,NewmanZiff01,FengDengBlote08}, from critical polynomials and transfer matrices~\cite{ScullardJacobsen12,JacobsenScullard13,Jacobsen14,Jacobsen15,ScullardJacobsen20}, or within rigorous confidence intervals~\cite{RiordanWalters07}.

The tilings of the plane by regular polygons are a natural family of test lattices. A tiling is $k$-uniform if it is edge-to-edge and its vertices form $k$ orbits under its symmetry group~\cite{GrunbaumShephard87}. There are 11 one-uniform (Archimedean) tilings, 20 two-uniform tilings~\cite{GrunbaumShephard87}, 61 three-uniform tilings~\cite{Chavey84,Chavey89}, and the number grows roughly as $2^k$, to 11\,866 at $k=10$; the enumeration is due to Galebach~\cite{Galebach,OEIS_A068599} and was extended by \v{C}trn\'act~\cite{OEIS_A068599,TesCatalog}. Of the 21 possible vertex types, 15 occur in tilings~\cite{GrunbaumShephard87}. Tilings by regular polygons appear as lattices of real materials~\cite{Li22}, as periodic approximants of dodecagonal square--triangle tilings~\cite{OxborrowHenley93}, and as test lattices for spin models~\cite{Yu17}; recently, the meshes of all $k$-uniform tilings with $k\le6$ were edge-colored to construct quantum circuits~\cite{Kattemolle25}. Thresholds of aperiodic tilings are known for the Penrose tiling and for octagonal and dodecagonal quasilattices~\cite{Babalievski95,ZiffBabalievski99}.

The site and bond thresholds of the Archimedean lattices are known to high precision~\cite{SudingZiff99,Parviainen07,FengDengBlote08,Jacobsen14,Jacobsen15,ScullardJacobsen20}. For the 20 two-uniform lattices, Neher, Mecke, and Wagner obtained the site thresholds with an uncertainty of $2\times10^{-4}$~\cite{Neher08,Neher08supp}; their bond thresholds were computed with an uncertainty of about $4\times10^{-7}$ by Gu and Ziff but not published~\cite{GuZiff}, and four-decimal values are listed by Lebrecht \textit{et al.}~\cite{Lebrecht21}. We did not find any published thresholds for $k\ge3$. The many proposed ``universal'' formulas for $\pc$~\cite{ScherZallen70,Vyssotsky61,GalamMauger96,GalamMauger97,GalamMauger98,vanderMarck97,WiermanNaor05,Neher08,Lebrecht21} have been calibrated and tested on a few dozen two-dimensional lattices at most.

In this paper we compute both thresholds of all 23\,396 $k$-uniform tilings with $k\le10$. The 486 thresholds with $k\le4$ have uncertainties of a few units in the sixth decimal, the remaining 46\,306 have median statistical uncertainties of $2\times10^{-5}$ ($k\le7$) to $7\times10^{-5}$ ($k\ge8$). Most of these tilings have no rotation of order 3, 4, or 6, so their large-scale anisotropy, and with it the effective shape of a simulated torus, is unknown. We therefore locate the threshold by the equality of the probabilities of no wrapping and of a cross-wrapping cluster. This is the Monte Carlo form of the critical-polynomial criterion~\cite{JacobsenScullard13}, and it has been used before for other models~\cite{Xu21,Bell22}. It needs no knowledge of the anisotropy (Sec.~\ref{sec:method}), the same property that makes critical polynomials applicable to arbitrary periodic lattices; our contribution is its systematic application and validation on a complete class of low-symmetry lattices. The data set allows a quantitative look at two old questions: how well $\pc$ can be predicted from local information, and how strongly it depends on the arrangement of a given set of vertex types. We find that rules based on the vertex composition alone predict $\pc$ to about $2\times10^{-3}$, while the arrangement shifts $\pc$ by up to $2\times10^{-2}$ (bond; $6\times10^{-3}$ for site), and that, for the domain sizes that occur at $k\le10$, descriptors of the neighborhood of radius one capture more than half of this shift. The smallest arrangement change, a slip of strips by half a period, gives threshold differences that decay exponentially with the distance between the strips, at a rate comparable to a transfer-matrix estimate.

The paper is organized as follows. Section~\ref{sec:tilings} describes the conversion of the tiling catalog into periodic graphs. Section~\ref{sec:method} describes the simulations, the threshold estimator, and its validation. Section~\ref{sec:results} presents the thresholds and tests empirical rules. Section~\ref{sec:slip} analyzes the slip pairs, and Sec.~\ref{sec:concl} concludes. Appendix~\ref{app:tm} gives the transfer-matrix details and Appendix~\ref{app:tab} lists the thresholds of the 61 three-uniform tilings. The thresholds of all tilings are given in the Supplemental Material~\cite{SM}.

\section{Tilings and periodic graphs}\label{sec:tilings}

\begin{figure*}
\includegraphics[width=\textwidth]{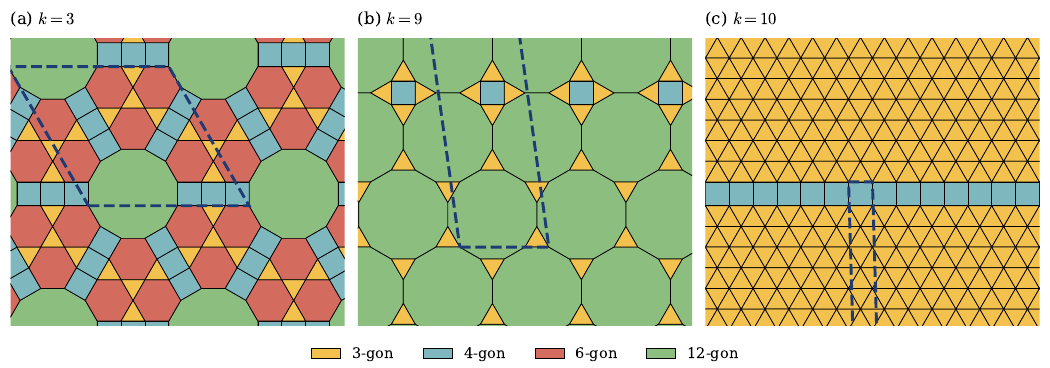}
\caption{Examples of $k$-uniform tilings; the dashed parallelograms are unit cells. (a) A three-uniform tiling with vertex types $\tfrac25(3.4^2.6)+\tfrac25(4.6.12)+\tfrac15(3.6.3.6)$, $\ps=0.6857748(35)$, $\pb=0.5911238(28)$. (b) The tiling with the highest site and bond thresholds among the non-Archimedean tilings with $k\le10$, $\tfrac{11}{13}(3.12^2)+\tfrac{2}{13}(3.4.3.12)$ ($k=9$), $\ps=0.79687(5)$, $\pb=0.72270(6)$. (c) The tiling with the lowest non-Archimedean thresholds, $\tfrac{9}{10}(3^6)+\tfrac{1}{10}(3^3.4^2)$ ($k=10$), $\ps=0.50535(7)$, $\pb=0.35441(6)$: one row of squares in every twenty rows of polygons.}
\label{fig:examples}
\end{figure*}

We take the tilings from the tessellation catalog of \v{C}trn\'act and Kopczy\'nski~\cite{TesCatalog}, which contains, among other tilings, all $k$-uniform tilings with $k\le10$. The numbers of tilings, 11, 20, 61, 151, 332, 673, 1472, 2850, 5960, and 11\,866 for $k=1,\dots,10$ (Table~\ref{tab:counts}), agree with the known enumeration~\cite{OEIS_A068599}. Up to $k=7$ this is the independent enumeration of Galebach~\cite{Galebach}; the numbers for $k=8$--10 were found by \v{C}trn\'act, a coauthor of the catalog~\cite{OEIS_A068599}, so for $k\ge8$ the agreement is a consistency check, and the completeness of our set rests on that enumeration. Each catalog file specifies the polygons, the gluing of their edges, and the vertex type of every vertex orbit.

From each file we construct a periodic graph. A patch of polygons is laid out by breadth-first gluing; the translation lattice is found by testing, in order of increasing length, the vectors from a vertex to the other vertices of the same type, a vector being accepted if it maps all vertices and edge midpoints within a disk of radius at least $1.1(|\mathbf T_1|+|\mathbf T_2|)+5$ edge lengths onto vertices and edge midpoints of the patch (the patch is enlarged until this holds); the basis is Lagrange--Gauss reduced to $(\mathbf T_1,\mathbf T_2)$ with $|\mathbf T_1|\le|\mathbf T_2|$; and the vertices are identified modulo translations. The result is a cell with $n_v$ vertices and a list of $n_e$ edges $(u,w,a,b)$, each joining vertex $u$ of cell $(0,0)$ to vertex $w$ of cell $(a,b)$.

We then verified the periodic graphs directly, independently of the patch. For every tiling, all edges have unit length; the angles between consecutive edges at every vertex reproduce the vertex type of the catalog; and every face, traced through the cyclic order of the edges, closes after $n$ steps without net translation and is an $n$-gon at each of its corners. The cell thus describes an edge-to-edge tiling by regular polygons, with $n_v-n_e+n_f=0$. No vector other than a lattice vector maps the vertices and edges of the cell onto vertices and edges modulo translations, so the cell is primitive. We computed the full symmetry group of every periodic graph exactly: an isometry is fixed by the images of one vertex and one of its edges and by its orientation, and a candidate is a symmetry if its linear part maps the translation lattice onto itself and it maps the vertices and edges of the cell onto vertices and edges modulo translations. For all 23\,396 tilings the vertices form exactly $k$ orbits. In the same way, two tilings are identical if and only if an isometry maps one periodic graph onto the other; among the 120\,597 pairs of tilings with equal composition, number of vertices per cell, and cell area (invariants of an isometry), there is no such pair. The 23\,396 periodic graphs are therefore pairwise distinct $k$-uniform tilings, and since their numbers agree with the enumeration, they are all $k$-uniform tilings with $k\le10$, as far as the enumeration is complete. Figure~\ref{fig:examples} shows three examples.

We label vertex types by the cyclic sequence of polygons around a vertex~\cite{GrunbaumShephard87}, e.g., $(3.4^2.6)$ for triangle, square, square, hexagon, and a tiling by its composition, the fractions of vertices of each type; for instance $\tfrac45(3.4^2.6)+\tfrac15(3.6.3.6)$ is the composition of two different two-uniform tilings. Of the 243 tilings with $k\le4$, 147 have no rotation of order 3, 4, or 6; such lattices need not be isotropic at large scales. The share of such tilings grows with $k$, to 94\% at $k=10$; of all 23\,396 tilings, 21\,319 (91\%) have no such rotation. The unit cells reach $n_v=120$ vertices at $k=10$.

\begin{table}
\caption{Number of $k$-uniform tilings, the protocol, and the median uncertainty of the site and bond thresholds (in units of $10^{-6}$). For $k\le4$: error of the finite-size extrapolation (statistical); changing the correction exponent shifts the values by a median of $0.1\sigma$. For $k\ge5$: statistical error; a direct check bounds the finite-size bias at $0.23\sigma$ for typical tilings and $0.30\sigma$ for the largest cells (95\% bounds, Sec.~\ref{sec:method}).}
\label{tab:counts}
\begin{ruledtabular}
\begin{tabular}{rrlrr}
$k$ & tilings & tori & $\sigma_{\rm site}$ & $\sigma_{\rm bond}$\\
\hline
1 & 11 & 3 tori, extrap. & 3.3 & 2.6 \\
2 & 20 & 3 tori, extrap. & 3.5 & 2.6 \\
3 & 61 & 3 tori, extrap. & 3.5 & 2.7 \\
4 & 151 & 3 tori, extrap. & 2.9 & 2.4 \\
5 & 332 & 1 torus & 21.4 & 18.4 \\
6 & 673 & 1 torus & 21.4 & 18.5 \\
7 & 1472 & 1 torus & 22.9 & 19.9 \\
8 & 2850 & 1 torus & 63.0 & 55.6 \\
9 & 5960 & 1 torus & 64.4 & 57.9 \\
10 & 11866 & 1 torus & 74.9 & 68.6 \\

\end{tabular}
\end{ruledtabular}
\end{table}

\section{Method}\label{sec:method}

\subsection{Simulation}

We use the microcanonical algorithm of Newman and Ziff~\cite{NewmanZiff00,NewmanZiff01} on tori of $L_x\times L_y$ unit cells. In each sample the $N$ elements, sites or bonds, of the torus are occupied one by one in a uniformly random order, and clusters are maintained by union--find with path compression. Every element stores its displacement to its parent, so that when an occupied element closes a cycle within a cluster, the winding vector of the cycle is obtained at no extra cost~\cite{NewmanZiff01}. For each sample we record the number of occupied elements $n_e$ at which a cluster first wraps the torus in any direction, and the number $n_x$ at which a single cluster first acquires two independent winding vectors (a cross configuration); the sample stops at $n_x$. The canonical probabilities follow by the exact binomial convolution,
\begin{equation}
R_X(p)=\frac1M\sum_{j=1}^{M}\Pr\!\left[\mathrm{Bin}(N,p)\ge n_X^{(j)}\right],\quad X=e,x,
\end{equation}
where $M$ is the number of samples. The statistical errors of $R_X(p)$ follow from the variance of the summands, so no binning is needed. Random numbers are generated by xoshiro256** (sequential code) or xoshiro128** (data-parallel code). Every run, and in the data-parallel code every thread and kernel launch, starts from a distinct 64-bit seed, expanded to the generator state by splitmix64; the seeds are derived from a hash of the run name and are listed with the job lists. We do not use the jump functions of the generators, so the streams are distinct but not provably disjoint; with periods of $2^{256}-1$ and $2^{128}-1$ and fewer than $10^{15}$ random numbers in total, the probability of an overlap is negligible. The two codes give identical histograms when run on identical random-number streams.

\subsection{Threshold estimator}

Let $P_0(p)=1-R_e(p)$ be the probability that no cluster wraps the torus and $P_+(p)=R_x(p)$ the probability of a cross configuration. For critical percolation in the scaling limit, Pinson showed that the two are equal, $\pi(\mathbb Z\times\mathbb Z)=\pi(0)$ in his notation, for every modular parameter $\tau$ of the torus~\cite{Pinson94}; Arguin gave the corresponding relation for the Potts model~\cite{Arguin02}; see also Refs.~\cite{DiFrancesco87,ZiffLorenzKleban99}. We checked the identity numerically with Pinson's expressions for square and rhombic tori. The assumption we need is universality up to a linear map: the scaling limit of the critical model on a torus is that of isotropic critical percolation on a torus with some effective modular parameter $\tau_{\rm eff}$, which may differ from the geometric one. Then $P_0-P_+\to\pi(0)-\pi(\mathbb Z\times\mathbb Z)=0$ at $\pc$, whatever $\tau_{\rm eff}$ is. Duality is needed only to obtain the identity itself: under planar duality (bond percolation) or matching (site percolation), a configuration without a wrapping cluster corresponds to a complementary configuration that contains a cross cluster, and vice versa, on every finite torus~\cite{MertensZiff16}, so that $P_0(p)=P_+^\ast(1-p)$ for the dual (matching) model, and for a self-dual or self-matching model $P_0=P_+$ at $\pc$. We therefore define the finite-size estimate $p_N$ as the root of
\begin{equation}
R_e(p)+R_x(p)=1.
\label{eq:cross}
\end{equation}
The complementarity of no-wrapping and cross configurations is an exact topological statement~\cite{MertensZiff16}. Universality is proven only for site percolation on the triangular lattice~\cite{Smirnov01} and is otherwise an assumption, which we test numerically below (Sec.~\ref{sec:valid}), including the anisotropic lattice $(3^3.4^2)$. Since every summand $\Pr[\mathrm{Bin}(N,p)\ge n]$ with $1\le n\le N$ is a strictly increasing function of $p$, so is $R_e+R_x$, so Eq.~\eqref{eq:cross} has exactly one root on every finite torus; we locate it by Brent's method on a bracket with a sign change.
Equation~\eqref{eq:cross} is the condition $P_B=0$ for the critical polynomial of Jacobsen and Scullard at $q=1$, $P_B(p)=P_{2D}(p)-P_0(p)$, whose roots, computed exactly on small tori, converge to $\pc$ very rapidly~\cite{JacobsenScullard13,Jacobsen14}; Mertens and Ziff derived the underlying topological identities for finite matching lattices from Euler's formula~\cite{MertensZiff16}. Monte Carlo estimates of the same root were used by Xu \textit{et al.} for nonplanar and continuum models, where they show small finite-size corrections~\cite{Xu21}, and, with Newman--Ziff sampling and binomial convolution as here, by Bell \textit{et al.} for the random-cluster model~\cite{Bell22}. What makes the criterion suited to the present lattices is that, in contrast to the common criterion $R_e(p)=R_e^\infty(\tau)$, it needs no knowledge of the universal value $R_e^\infty$: for a lattice without three-, four-, or sixfold rotations, $\tau_{\rm eff}$ may differ from the geometric $\tau$ by an unknown anisotropy, and $R_e^\infty(\tau_{\rm eff})$ is then not known. For the uncertainty of $p_N$ we use $(\sigma_e+\sigma_x)/|d(R_e+R_x)/dp|$, where $\sigma_X$ is the standard error of $R_X(p_N)$ and the derivative is the central finite difference of the exact convolution, with step $2\times10^{-3}/\sqrt N$; its own statistical error is negligible. This is an upper bound for any correlation of $R_e$ and $R_x$; the histograms do not record the joint distribution of $n_e$ and $n_x$, so the exact variance is not available. The independent comparisons below give rms deviations between 0.71 and 0.99 combined standard errors, 0.88 for the 4954 thresholds of the two atlas rounds, so the bound exceeds the actual error by about 10--20\%. Our $\chi^2$ tests are correspondingly conservative: with errors smaller by a factor 0.88, the $\chi^2$ values of the comparison in Sec.~\ref{sec:results} would increase from 19.5 to 25 for 20 values and from 6.0 to 7.8 for 9, still acceptable.

\subsection{Finite-size corrections}

For the estimator $R_e(p)=R_e^\infty$, the estimate $p_L$ on a torus of linear size $L$ approaches $\pc$ as $L^{-2-1/\nu}=L^{-11/4}$~\cite{NewmanZiff00,NewmanZiff01}, while the roots of the exactly computed critical polynomials converge empirically as $L^{-w}$ with $w\approx4$~\cite{MertensZiff16}. We use the slower law $L^{-11/4}$ and check below that the results are insensitive to this choice. For $k\le4$ (the ``core'') we simulated three tori per threshold, with $L_x|\mathbf T_1|\approx L_y|\mathbf T_2|$ (aspect ratios 0.87--1.15) and median sizes of 504, 2016, and 8100 sites, and fitted
\begin{equation}
p_N=\pc+aN_s^{-11/8},
\label{eq:fss}
\end{equation}
with $N_s$ the number of sites of the torus (also for bond percolation, where the number of bonds $N$ is proportional to $N_s$). The fits have one degree of freedom; the median $\chi^2$ is 0.37 and 2.7\% of the fits exceed the 95\% point 3.84. The quoted errors are the fit errors, not rescaled by $\chi^2$. With these sizes and exponent, the smallest torus mainly fixes the amplitude $a$: the extrapolated values differ from the weighted mean of the two larger tori by a median of $0.17\sigma$ (90th percentile $0.48\sigma$, maximum $1.9\sigma$); the torus values are given in the Supplemental Material~\cite{SM}. Changing the exponent from $11/4$ to $4$ changes $\pc$ by a median of $0.10\sigma$ and at most $0.81\sigma$; with exponents between 2 and 5 the changes stay below $1.2\sigma$ (median $\le0.14\sigma$). The median number of samples is $1.3\times10^8$ per threshold, $7.5\times10^{10}$ in total.

\begin{figure*}
\includegraphics[width=\textwidth]{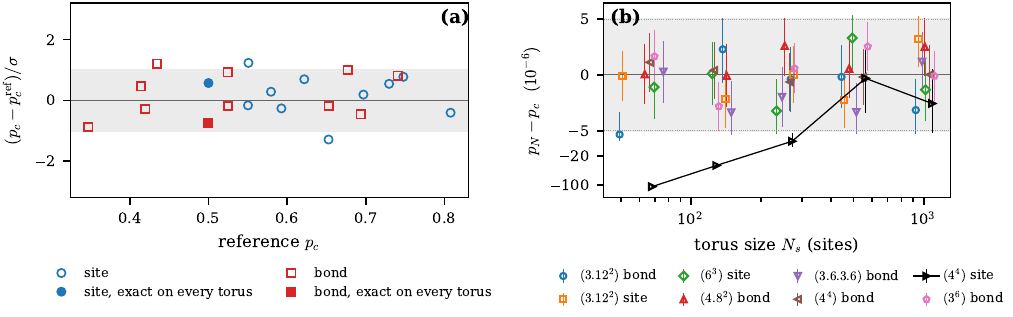}
\caption{(a) Deviation of our thresholds from the 22 Archimedean reference values (exact values and Refs.~\cite{SudingZiff99,Jacobsen14,Jacobsen15,ScullardJacobsen20}) in units of the combined standard error. Filled symbols: the two thresholds that are exact on every finite torus by self-matching or self-duality, $\ps(3^6)$ and $\pb(4^4)$; they test the statistical errors only. The shaded band is $\pm1\sigma$. (b) Finite-size estimates $p_N$ of Eq.~\eqref{eq:cross} minus the reference value for eight Archimedean thresholds, as a function of the torus size; the vertical axis is linear for $|p_N-\pc|<5\times10^{-6}$ and logarithmic beyond. Only the square-lattice site threshold (connected symbols) shows a visible correction.}
\label{fig:validation}
\end{figure*}

For $k=5$--10 (the ``atlas'') each threshold comes from a single torus with $N_s\ge738$ sites (median 1024 for $k=5$, 1764 for $k=10$) and at least three unit cells in each direction; since the edges of a cell join only neighboring cells, the torus graphs have no loops or multiple edges. For many tilings with long unit cells this torus is elongated: 16\% of the atlas tori have an aspect ratio below 1/2 (minimum 0.35). The runs for $k=5$--7 were made in two independent rounds on the same tori; the two rounds agree with an rms deviation of $0.88$ combined standard errors over 4954 thresholds, which confirms the statistical errors. In total the atlas used $1.4\times10^{10}$ samples. The atlas values carry no finite-size correction, and their quoted errors, $2\times10^{-5}$ ($k\le7$) and $(5.6$--$7.5)\times10^{-5}$ ($k\ge8$), are statistical. The amplitudes $a$ fitted in the core suggest that the residual bias is smaller: at $N_s=1000$, $|a|N_s^{-11/8}$ is below $4.5\times10^{-6}$ for 90\% and below $1.6\times10^{-5}$ for 99\% of the 486 core thresholds. Since the atlas contains larger cells ($n_v$ up to 120) and more elongated tori than the core, we checked the bias directly on a stratified sample of 130 atlas tilings: the 35 with the most elongated tori (all with aspect ratio 0.35--0.36), the 35 with the largest cells ($n_v=102$--120), 5 random tilings for each $k=5,\dots,10$, and, since the only visible finite-size correction in Fig.~\ref{fig:validation}(b) is that of the square lattice, the 30 further tilings with the largest fraction of $(4^4)$ vertices (0.79--0.90), i.e., the largest square domains. For each of these 260 thresholds we extended the run on the atlas torus and simulated a near-square torus with 3.7--4.6 times as many sites ($1.25\times10^9$ additional samples in total). The extension runs are independent of the original atlas runs; for the 216 thresholds with $k=8$--10, where the atlas has only one round, they agree with the atlas values with an rms deviation of 0.81 combined standard errors (no deviation above $3\sigma$), which confirms the statistical errors also for $k\ge8$. The difference $\Delta=p_{N_2}-p_{N_1}$ between the larger and the atlas torus measures the finite-size bias of the atlas value (up to a factor $1-(N_2/N_1)^{-\omega/2}\approx0.83$--0.95 for correction exponents $\omega$ between $11/4$ and 4); its uncertainty is about half the atlas error. We find $\langle\Delta/\sigma_\Delta\rangle=-0.03$ and an rms of 0.91, with no $|\Delta|>3\sigma_\Delta$; the largest $|\Delta|$ is 1.75 atlas standard errors. In the four strata the rms of $\Delta/\sigma_\Delta$ is 0.92 (random), 0.88 (elongated), 1.00 (largest cells), and 0.81 (square domains), with means between $-0.19$ and $+0.10$. We model the bias as Gaussian with an rms of $\beta$ atlas standard errors; a profile-likelihood fit gives $\beta$ consistent with zero in every stratum, with 95\% upper bounds of 0.23 for the random sample, which represents typical tilings, 0.21 for the most elongated tori, 0.30 for the largest cells, and 0.16 for the largest square domains. Within the precision of the atlas, its values are therefore free of finite-size bias. The check runs are listed in the Supplemental Material~\cite{SM}.

Figure~\ref{fig:validation}(b) shows the finite-size behavior directly for eight Archimedean thresholds, simulated on tori of 54 to 1058 sites with $\sigma\approx3\times10^{-6}$ each. For seven of them no deviation from the exact or high-precision value is visible even on the smallest tori. The site threshold of the square lattice is the exception: $p_N-\pc=-108$, $-34$, and $-9\times10^{-6}$ for $N_s=64$, 121, and 256, and zero within errors for $N_s\ge529$. This correction decreases about as $N_s^{-1.8}$, faster than assumed in Eq.~\eqref{eq:fss}, and extrapolates to $8\times10^{-7}$ at $N_s=1000$, so it is negligible at the atlas sizes also for tilings with large square domains, as the direct check of the square-domain stratum above confirms. The exponent, $N_s^{-1.8}\approx L^{-3.6}$, is close to the $w\approx4$ of the critical-polynomial roots~\cite{MertensZiff16}; what is unusual is the amplitude. In the core fits no finite-size amplitude is significant at the $3\sigma$ level.

\subsection{Validation}\label{sec:valid}

Table~\ref{tab:arch} and Fig.~\ref{fig:validation}(a) compare our values with the 22 Archimedean thresholds. The deviation is $+0.18\sigma$ on average and $0.71\sigma$ rms, with a maximum of $1.3\sigma$; without the two thresholds that are exact on every finite torus, the rms deviation is $0.72\sigma$. The error bars are thus realistic. The site threshold of $(3^2.4.3.4)$ was the only one that deviated by more than two standard errors in the core run, $0.5508171(38)$ against $0.550806(3)$ from Ref.~\cite{SudingZiff99}. A dedicated run on the same three tori, with about 40 times the statistics on the two smaller ones and four times on the largest, gives $0.5508098(7)$ (Table~\ref{tab:arch}), 1.2 combined standard errors from Ref.~\cite{SudingZiff99}; the core value was thus a fluctuation of $1.9\sigma$. Correction exponents between 2 and 5 change the dedicated value by at most $0.6\sigma$. No other threshold was rerun. With the core value, the rms deviation over the 22 references is $0.82\sigma$ and the maximum $2.3\sigma$; the dedicated value is used in all tables, in the Supplemental Material, and in the fits of Sec.~\ref{sec:results}. Our values for $(3^4.6)$ and $(3^3.4^2)$, the other two site thresholds known only to this precision, agree with Ref.~\cite{SudingZiff99}. The reference set contains no published high-precision threshold of a lattice with more than one vertex orbit; the unpublished high-precision bond thresholds of the two-uniform lattices~\cite{GuZiff} provide such a test (Sec.~\ref{sec:results}).

The code itself can be tested exactly on small tori. On a torus with $K$ elements, the first $m$ elements of a uniformly random order form a uniform $m$-subset, so the distributions of $n_e$ and $n_x$ follow from the numbers of $m$-subsets without a wrapping cluster and with a cross cluster, which we obtain by full enumeration for $K\le30$; for $K\le20$ an independent program, based on breadth-first search instead of union--find, gives the same numbers. For 12 tori of the square, triangular, honeycomb, kagome, and $(3^3.4^2)$ lattices and of a two- and a three-uniform tiling, with $K=16$--28 sites or bonds, we compared the distributions of $n_e$ and $n_x$ from $2\times10^9$ samples per torus with the exact ones: all 24 $\chi^2$ tests pass (smallest $p$-value 0.016). The roots of Eq.~\eqref{eq:cross} agree with the exact finite-torus roots with $\chi^2=6.1$ for 12 values, at a precision of $1.7\times10^{-6}$. This tests the sampling, the winding detection, the convolution, and the root finding without any extrapolation. For the triangular and honeycomb bond and the kagome site thresholds, which are known exactly, the exact roots on these small tori already equal $\pc$, as expected for these solvable lattices~\cite{JacobsenScullard13}.

Simple sanity checks hold as well. The bond thresholds of the mutually dual triangular and honeycomb lattices sum to $0.9999976(35)$, which is not independent of the comparison in Table~\ref{tab:arch}. Hammersley's inequality $\pb\le\ps$~\cite{Hammersley61} and the bound $\ps\le1-(1-\pb)^{\Delta-1}$, where $\Delta$ is the maximal degree~\cite{GrimmettStacey98}, hold for all 23\,396 tilings, with margins of at least 0.044 and 0.12.

\begin{table*}
\caption{Archimedean lattices: our thresholds, the reference values, and the deviation $z=(\pc-p_c^{\rm ref})/\sigma$ in units of the combined standard error. Reference values written with ``\ldots'' are exact or known to more digits than shown; $\dagger$: exact on every finite torus.}
\label{tab:arch}
\begin{ruledtabular}
\begin{tabular}{lllrllr}
lattice & $\ps$ (this work) & $\ps$ (reference) & $z$ & $\pb$ (this work) & $\pb$ (reference) & $z$\\
\hline
$(3.12^{2})$ & 0.8078999(22) & 0.807900764\ldots~\cite{SudingZiff99} & -0.4 & 0.7404231(29) & 0.7404207989\ldots~\cite{ScullardJacobsen20} & +0.8 \\
$(4.6.12)$ & 0.7478030(28) & 0.7478008(2)~\cite{Jacobsen14} & +0.8 & 0.6937319(26) & 0.6937331249\ldots~\cite{ScullardJacobsen20} & -0.5 \\
$(4.8^{2})$ & 0.7297248(29) & 0.7297232(5)~\cite{Jacobsen14} & +0.5 & 0.6768058(26) & 0.6768031244\ldots~\cite{ScullardJacobsen20} & +1.0 \\
$(6^{3})$ & 0.6970408(30) & 0.697040230(5)~\cite{Jacobsen14} & +0.2 & 0.6527031(27) & 0.652703645\ldots~\cite{SykesEssam64} & -0.2 \\
$(3.6.3.6)$ & 0.6526994(33) & 0.652703645\ldots~\cite{SykesEssam64} & -1.3 & 0.5244075(27) & 0.5244049992\ldots~\cite{ScullardJacobsen20} & +0.9 \\
$(3.4.6.4)$ & 0.6218145(35) & 0.62181207(7)~\cite{Jacobsen14} & +0.7 & 0.5248310(27) & 0.5248314616\ldots~\cite{ScullardJacobsen20} & -0.2 \\
$(4^{4})$ & 0.5927452(33) & 0.5927460508\ldots~\cite{Jacobsen15} & -0.3 & 0.4999981(26) & 1/2~\cite{Kesten80}$^{\dagger}$ & -0.7 \\
$(3^{4}.6)$ & 0.5794994(38) & 0.579498(3)~\cite{SudingZiff99} & +0.3 & 0.4343316(27) & 0.4343283172\ldots~\cite{ScullardJacobsen20} & +1.2 \\
$(3^{2}.4.3.4)$ & 0.5508098(7) & 0.550806(3)~\cite{SudingZiff99} & +1.2 & 0.4141390(24) & 0.4141378566\ldots~\cite{ScullardJacobsen20} & +0.5 \\
$(3^{3}.4^{2})$ & 0.5502122(38) & 0.550213(3)~\cite{SudingZiff99} & -0.2 & 0.4196396(25) & 0.4196403589\ldots~\cite{ScullardJacobsen20} & -0.3 \\
$(3^{6})$ & 0.5000021(36) & 1/2~\cite{SykesEssam64}$^{\dagger}$ & +0.6 & 0.3472945(22) & 0.347296355\ldots~\cite{SykesEssam64} & -0.9 \\

\end{tabular}
\end{ruledtabular}
\end{table*}

\section{Results}\label{sec:results}

\subsection{Two-, three-, and four-uniform tilings}

\begin{table*}
\caption{Two-uniform lattices, ordered by decreasing site threshold. ID: catalog file name~\cite{TesCatalog}. Site thresholds are compared with Neher \textit{et al.}~\cite{Neher08supp} and bond thresholds with the unpublished values of Gu and Ziff~\cite{GuZiff}. $^\ast$The two lattices with composition $\tfrac45(3.4^2.6)+\tfrac15(3.6.3.6)$ differ in the registry of their strips (Sec.~\ref{sec:slip}); Ref.~\cite{Neher08supp} lists two values for this composition without identifying the lattices, and the assignment shown follows the pairing of these values with bond values in Ref.~\cite{Lebrecht21}. The bond values of Ref.~\cite{GuZiff} are matched to the two tilings by value.}
\label{tab:2uni}
\begin{ruledtabular}
\begin{tabular}{llllll}
ID & vertex types & $\ps$ (this work) & $\ps$~\cite{Neher08supp} & $\pb$ (this work) & $\pb$~\cite{GuZiff}\\
\hline
\texttt{3a 4b 1} & $\tfrac{1}{2}(3.12^{2})+\tfrac{1}{2}(3.4.3.12)$ & 0.7671769(30) & 0.7680(2) & 0.6749333(29) & 0.67493252(36) \\
\texttt{3c 4l 1} & $\tfrac{2}{3}(4.6.12)+\tfrac{1}{3}(3.4.6.4)$ & 0.7154422(31) & 0.7157(2) & 0.6453683(25) & 0.64536587(40) \\
\texttt{4a 6d 1} & $\tfrac{6}{7}(3^{2}.4.12)+\tfrac{1}{7}(3^{6})$ & 0.6806338(34) & 0.6808(2) & 0.5577774(29) & 0.55778329(40) \\
\texttt{4d 4f 1} & $\tfrac{2}{3}(3^{2}.6^{2})+\tfrac{1}{3}(3.6.3.6)$ & 0.6496372(35) & 0.6499(2) & 0.5363209(27) & 0.53632487(40) \\
\texttt{4d 6a 1} & $\tfrac{6}{7}(3^{2}.6^{2})+\tfrac{1}{7}(3^{6})$ & 0.6332093(44) & 0.6329(2) & 0.5170775(29) & 0.51707873(70) \\
\texttt{4f 4k 2 (eclipsed)} & $\tfrac{4}{5}(3.4^{2}.6)+\tfrac{1}{5}(3.6.3.6)$ & 0.6283846(37) & 0.6286(2)$^{\ast}$ & 0.5189135(26) & 0.51891529(35) \\
\texttt{4f 4k 1 (staggered)} & $\tfrac{4}{5}(3.4^{2}.6)+\tfrac{1}{5}(3.6.3.6)$ & 0.6281115(39) & 0.6279(2)$^{\ast}$ & 0.5176952(27) & 0.51769462(35) \\
\texttt{4k 4l 1} & $\tfrac{2}{3}(3.4^{2}.6)+\tfrac{1}{3}(3.4.6.4)$ & 0.6218435(33) & 0.6221(2) & 0.5197407(26) & 0.51973831(40) \\
\texttt{4d 5a 1} & $\tfrac{1}{2}(3^{4}.6)+\tfrac{1}{2}(3^{2}.6^{2})$ & 0.6169704(45) & 0.6171(2) & 0.4892140(29) & 0.48921280(37) \\
\texttt{4l 5e 1} & $\tfrac{1}{2}(3^{2}.4.3.4)+\tfrac{1}{2}(3.4.6.4)$ & 0.5881281(37) & 0.5883(2) & 0.4657312(26) & 0.46573078(72) \\
\texttt{4l 5c 1} & $\tfrac{1}{2}(3^{3}.4^{2})+\tfrac{1}{2}(3.4.6.4)$ & 0.5880904(35) & 0.5885(2) & 0.4722908(27) & 0.47229486(38) \\
\texttt{4s 5c 1} & $\tfrac{1}{2}(3^{3}.4^{2})+\tfrac{1}{2}(4^{4})$ & 0.5717644(37) & 0.5720(2) & 0.4584432(28) & 0.45844622(40) \\
\texttt{4p 5c 1} & $\tfrac{2}{3}(3^{3}.4^{2})+\tfrac{1}{3}(4^{4})$ & 0.5646310(36) & 0.5648(2) & 0.4452838(25) & 0.44528611(40) \\
\texttt{5a 6d 1} & $\tfrac{3}{4}(3^{4}.6)+\tfrac{1}{4}(3^{6})$ & 0.5605459(36) & 0.5607(2) & 0.4111037(26) & 0.41109890(37) \\
\texttt{5c 5f 1} & $\tfrac{2}{3}(3^{2}.4.3.4)+\tfrac{1}{3}(3^{3}.4^{2})$ & 0.5505125(30) & 0.5504(2) & 0.4154916(22) & 0.41549285(36) \\
\texttt{5d 5f 1} & $\tfrac{1}{2}(3^{3}.4^{2})+\tfrac{1}{2}(3^{2}.4.3.4)$ & 0.5504255(31) & 0.5505(2) & 0.4162784(28) & 0.41628021(35) \\
\texttt{5e 6a 1} & $\tfrac{6}{7}(3^{2}.4.3.4)+\tfrac{1}{7}(3^{6})$ & 0.5440521(35) & 0.5440(2) & 0.4037930(25) & 0.40379585(40) \\
\texttt{5b 6j 1} & $\tfrac{1}{2}(3^{6})+\tfrac{1}{2}(3^{4}.6)$ & 0.5408655(34) & 0.5407(2) & 0.3891478(21) & 0.38914898(35) \\
\texttt{5c 6f 1} & $\tfrac{2}{3}(3^{3}.4^{2})+\tfrac{1}{3}(3^{6})$ & 0.5342744(34) & 0.5342(2) & 0.3949216(26) & 0.39491996(40) \\
\texttt{5c 6i 1} & $\tfrac{1}{2}(3^{6})+\tfrac{1}{2}(3^{3}.4^{2})$ & 0.5260067(28) & 0.5258(2) & 0.3828495(20) & 0.38285085(38) \\

\end{tabular}
\end{ruledtabular}
\end{table*}

Table~\ref{tab:2uni} lists the two-uniform thresholds. For the 18 lattices that can be matched unambiguously, the site thresholds of Neher \textit{et al.}~\cite{Neher08supp} deviate from ours by $2.8\times10^{-4}$ rms; 10 of the 18 values lie within the quoted uncertainty of $2\times10^{-4}$ and 16 within twice that. The deviations are not random: they grow with $\pc$ (correlation coefficient 0.67), and after subtraction of a linear trend in $\pc$ their scatter is $1.9\times10^{-4}$, consistent with the quoted uncertainty. The values of Ref.~\cite{Neher08supp} thus appear to carry a small systematic error that grows with $\pc$. It does not depend on whether the lattice has rotations of order 3, 4, or 6: after the trend is subtracted, the mean residuals are $-0.2\times10^{-4}$ for the 11 isotropic and $+0.3\times10^{-4}$ for the 7 other lattices. A dependence on the shape of their simulation domains, which we cannot reconstruct, is possible. Our uncertainties are smaller by a factor of about 60. The largest deviation is for $\tfrac12(3.12^2)+\tfrac12(3.4.3.12)$, the only two-uniform tiling with this composition: we find $0.7671769(30)$ instead of $0.7680(2)$, a difference of $4.1$ quoted standard errors, or 2.3 relative to the trend of the other 17 lattices. Two checks support our value. The bond threshold of the same periodic graph agrees with the value of Ref.~\cite{Lebrecht21} (below). And since this tiling has fourfold rotations and a square unit cell, its square tori have $\tau=i$, so the common estimator $R_e(p)=R_e^\infty(i)=0.690473725$~\cite{Pinson94,NewmanZiff01} applies; extrapolated over the same three tori, it gives $0.7671739(30)$. The two estimates share the samples but have different finite-size corrections, and they agree.

The two tilings with composition $\tfrac45(3.4^2.6)+\tfrac15(3.6.3.6)$ have site thresholds $0.6283846(37)$ (eclipsed) and $0.6281115(39)$ (staggered); their difference, $2.73(5)\times10^{-4}$, is smaller than the difference $7(3)\times10^{-4}$ of the two values of Ref.~\cite{Neher08supp}. Bond thresholds of all 20 two-uniform lattices, with uncertainties of $(3.5$--$7)\times10^{-7}$, were computed by Gu and Ziff but not published~\cite{GuZiff}; Lebrecht \textit{et al.} list them rounded to four decimals~\cite{Lebrecht21}. Our values agree with them (Table~\ref{tab:2uni}): $\chi^2=19.5$ for 20 degrees of freedom, a weighted mean difference of $-0.8(6)\times10^{-6}$, and a largest deviation of 1.97 combined standard errors. For the 9 lattices without rotations of order 3, 4, or 6, where the effective torus shape is unknown, $\chi^2=6.0$ for 9 degrees of freedom. This tests the estimator at the level of $10^{-6}$ on lattices with two vertex orbits and confirms our error estimates there. Reference~\cite{Lebrecht21} pairs the two site values of Ref.~\cite{Neher08supp} for the composition $\tfrac45(3.4^2.6)+\tfrac15(3.6.3.6)$ with bond values, 0.6286 with 0.5189 and 0.6279 with 0.5177. Our bond thresholds of the eclipsed and staggered tilings, $0.5189135(26)$ and $0.5176952(27)$, differ by $1.2\times10^{-3}$, so this pairing fixes the assignment of the site values in Table~\ref{tab:2uni}; the assignment thus rests on Ref.~\cite{Lebrecht21}. To our knowledge, bond thresholds of the two-uniform lattices with error estimates have not been published.

The thresholds of the 61 three-uniform tilings are given in Tables~\ref{tab:3uni} and \ref{tab:3unib} (Appendix~\ref{app:tab}); those of the 151 four-uniform tilings are in the Supplemental Material~\cite{SM}. The site thresholds of the three-uniform tilings range from 0.517541 to 0.729771 and the bond thresholds from 0.370944 to 0.633689; for $k=4$ the ranges are 0.513227--0.787193 and 0.365030--0.707756.

Through exact duality and matching relations, the thresholds also give those of related lattices: $1-\pb$ is the bond threshold of the dual tiling and $1-\ps$ the site threshold of the matching graph, obtained by joining all vertices of every face. For site percolation the matching relation $\ps(G)+\ps(G^\ast)=1$ is a theorem for all amenable quasi-transitive planar graphs~\cite{GrimmettLi24hyperbolic}, hence for all our tilings. For bond percolation the duality relation $\pb(G)+\pb(G^+)=1$, where $G^+$ is the dual graph, holds for the same class of graphs~\cite{LyonsPeres16}, as noted in Ref.~\cite{GrimmettLi24hyperbolic} (Remark~1.5); it therefore does not depend on the reflection symmetry assumed in the classical proof~\cite{Kesten82,WiermanZiff11}, which many of our tilings lack.

\subsection{All tilings with $k\le10$}

\begin{figure*}
\includegraphics[width=\textwidth]{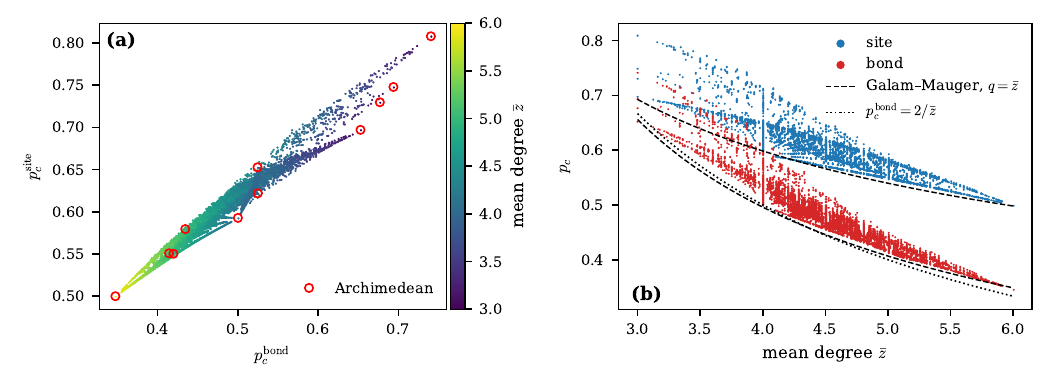}
\caption{(a) Site versus bond threshold for all 23\,396 tilings with $k\le10$, colored by the mean degree $\zb$; circles mark the Archimedean lattices. (b) Site (blue) and bond (red) thresholds versus $\zb$, with the Galam--Mauger formula~\cite{GalamMauger96} evaluated at $q=\zb$ (dashed) and the rule $\pb=2/\zb$~\cite{Vyssotsky61} (dotted). The vertical streaks, e.g., at $\zb=4$ and 5, are tilings with a single vertex degree.}
\label{fig:atlas}
\end{figure*}

Figure~\ref{fig:atlas} gives an overview of all 46\,792 thresholds. All of them lie between the Archimedean extremes: $\tfrac12\le\ps\le0.807901$, attained by $(3^6)$ and $(3.12^2)$, and $0.347296\le\pb\le0.740421$, attained by the same two lattices. The lower bound on $\ps$ is rigorous: the matching graph $G^\ast$ contains the tiling $G$, so $\ps(G^\ast)\le\ps(G)$, and $\ps(G)+\ps(G^\ast)=1$~\cite{GrimmettLi24hyperbolic}. It is strict for every tiling except $(3^6)$. For quasi-transitive plane graphs, $p_u+\pc\ge1$ for site percolation, with equality if and only if the graph obtained by emptying all separating triangles is a triangulation~\cite{Grimmett25} (for transitive graphs, Ref.~\cite{GrimmettLi24matching}), and $p_u=\pc$ for our amenable graphs~\cite{GrimmettLi24hyperbolic}. In a tiling by regular polygons every 3-cycle bounds a triangular face, since its interior is too small to contain a vertex, so there are no separating triangles, and the graph is a triangulation only for $(3^6)$. Numerically, $\ps-\tfrac12\ge5.3\times10^{-3}$ (72 standard errors) for every other tiling. The non-Archimedean tilings closest to the bounds are shown in Fig.~\ref{fig:examples}: the triangular lattice with one row of squares in every twenty rows, $\ps=0.50535(7)$, and the $(3.12^2)$ lattice interrupted by rows of $(3.4.3.12)$ vertices, $\ps=0.79687(5)$. The same two tilings also have the extreme non-Archimedean bond thresholds. The site and bond thresholds are strongly correlated (correlation coefficient 0.978), but at a fixed bond threshold the site threshold varies by up to 0.05, as the band structure of Fig.~\ref{fig:atlas}(a) shows.

\subsection{Empirical rules}

\begin{table*}
\caption{Accuracy of empirical rules: rms of the estimate minus the threshold on the hold-out tilings with $k\ge5$, with every tiling weighted equally. ``A'': the 6504 tilings whose vertex types all occur in Archimedean lattices, where Eq.~\eqref{eq:avg} is defined; ``all'': all 23\,153 tilings with $k\ge5$. The column ``fitted'' gives the number of parameters fitted on the 243 tilings with $k\le4$ (zero for rules taken as published). The last two rows predict $\ps$ from the measured $\pb$.}
\label{tab:rules}
\begin{ruledtabular}
\begin{tabular}{lrllll}
 & & \multicolumn{2}{c}{site} & \multicolumn{2}{c}{bond}\\
rule & fitted & A & all & A & all\\
\hline
Galam--Mauger~\cite{GalamMauger96}, $q=\zb$ & 0 & 0.018 & 0.033 & 0.016 & 0.028 \\
Galam--Mauger form, refitted & 2 & 0.014 & 0.016 & 0.0075 & 0.013 \\
Euler estimate $p^\ast$~\cite{Neher08} & 0 & 0.0028 & 0.0048 & 0.0074 & 0.016 \\
$p^\ast$ + constant, refitted & 1 & 0.0048 & 0.0048 & 0.0097 & 0.011 \\
linear in $f_n$ (and $\zb$), refitted & 5 & 0.0026 & 0.0028 & 0.0030 & 0.0046 \\
Archimedean average, Eq.~\eqref{eq:avg} & 0 & 0.0015 & --- & 0.0017 & --- \\
linear rule, Eq.~\eqref{eq:mix} & 15 & 0.0010 & 0.0019 & 0.0010 & 0.0023 \\
linear rule for $\ln[\pc/(1-\pc)]$ & 15 & 0.0007 & 0.0014 & 0.0009 & 0.0021 \\
\hline
Lebrecht \textit{et al.}~\cite{Lebrecht21}, $\ps$ from $\pb$ & 0 & 0.014 & 0.019 & & \\
Galam--Mauger~\cite{GalamMauger98}, $\ps$ from $\pb$ & 0 & 0.0091 & 0.016 & & \\

\end{tabular}
\end{ruledtabular}
\end{table*}

\begin{figure*}
\includegraphics[width=\textwidth]{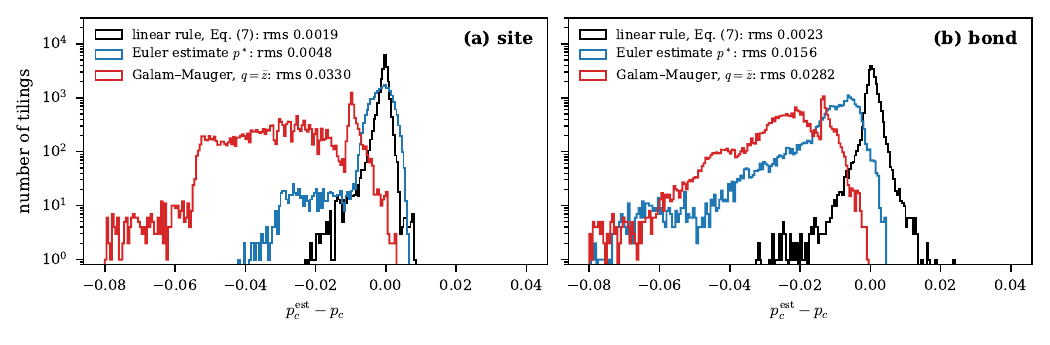}
\caption{Distribution of the errors of empirical estimates over the 23\,153 hold-out tilings with $k\ge5$: (a) site, (b) bond. Black: linear rule, Eq.~\eqref{eq:mix}, fitted on $k\le4$; blue: Euler-characteristic estimate $p^\ast$ of Ref.~\cite{Neher08}; red: Galam--Mauger formula~\cite{GalamMauger96} at $q=\zb$ (a few values below $-0.08$ are not shown).}
\label{fig:formulas}
\end{figure*}

The data set is large enough to test empirical rules for $\pc$ in a regime where they were not calibrated. To compare rules that use different information on equal terms, we score all of them on the same hold-out sets of tilings with $k\ge5$; rules with free parameters are fitted on the 243 tilings with $k\le4$ only (Table~\ref{tab:rules}, Fig.~\ref{fig:formulas}). The fits are unweighted, and the hold-out tilings are mostly mixtures of common vertex types, which makes the hold-out set somewhat easier than the training set.

\textit{Invariants.} Scher and Zallen observed that the critical filling fraction of disks centered on the occupied sites, with radius half the nearest-neighbor distance, is nearly the same for the common lattices: $0.44\pm0.02$ for the four two-dimensional lattices they considered~\cite{ScherZallen70}; Suding and Ziff found that a generalized filling factor correlates the Archimedean site thresholds well~\cite{SudingZiff99}. With unit edges and disks of diameter one, $\phi_c=\ps\rho\pi/4$, where $\rho$ is the number of vertices per unit area. Over all 23\,396 tilings we find $\phi_c=0.453\pm0.012$ (mean and standard deviation; 90\% of the values between 0.440 and 0.464). The invariant fails for open tilings with 8- and 12-gons, down to $\phi_c=0.316$ for $(3.12^2)$; without the 889 tilings that contain such polygons, $\phi_c=0.455\pm0.006$. For the bond rule $\zb\pb\approx d/(d-1)=2$, which Vyssotsky \textit{et al.} proposed tentatively~\cite{Vyssotsky61}, we find $\zb\pb=2.119\pm0.053$: the product is as constant as $\phi_c$ (relative standard deviation 2.5\%), but its value is 6\% above 2.

\textit{Universal formulas.} The formula of Galam and Mauger, $\pc=p_0[(d-1)(q-1)]^{-a}d^{\,b}$ with the two-dimensional parameters $p_0=0.8889$, $a=0.3601$, $b=0$ (site) and $p_0=0.6558$, $a=b=0.6897$ (bond)~\cite{GalamMauger96}, evaluated at $q=\zb$, underestimates the thresholds by 0.03 on average [Fig.~\ref{fig:atlas}(b)]; refitting $p_0$ and $a$ halves the error. Galam and Mauger noted that non-Bravais lattices require an effective coordination number~\cite{GalamMauger97}, which is not known in advance. Two rules predict the site threshold from the bond threshold. The ratio formula $\ps/\pb\approx a\zb^3/(\zb-1)^2$ with $a=0.16428$ of Lebrecht \textit{et al.}~\cite{Lebrecht21} was fitted to the 31 Archimedean and two-uniform lattices, with an average relative error of 2.7\%. Its average relative error is 2.6\% on those 31 lattices with our thresholds and 2.9\% on all tilings, but the residual is systematic: the formula underestimates the ratio by $0.037$ on average. The site--bond invariant $(\ps)^{1/a_s}(\pb)^{-1/a_b}=\delta/d$ of Galam and Mauger~\cite{GalamMauger98}, with $a_s=0.3670$, $\delta=1.3638$, and $a_b=0.6897$, does somewhat better (Table~\ref{tab:rules}).

\textit{Euler characteristic.} Neher \textit{et al.}~\cite{Neher08} showed that the mean Euler characteristic per site $\chi(p)$ of the occupied cluster pattern is a polynomial determined by the local structure. For site percolation on a tiling with mean degree $\zb$ and $f_n$ $n$-gons per vertex,
\begin{equation}
\chi(p)=p-\frac{\zb}{2}p^2+\sum_n f_np^n,
\end{equation}
and for bond percolation, per bond, on the covering lattice,
\begin{equation}
\chi(p)=\frac{2}{\zb}\Bigl[\sum_v x_v\bigl(1-(1-p)^{z_v}\bigr)+\sum_nf_np^n\Bigr]-p,
\end{equation}
where $x_v$ is the fraction of vertices of degree $z_v$. Neher \textit{et al.} proposed the estimate $p^\ast$ given by $\chi(p^\ast)=\frac{\tau-2}{\tau-1}(2p^\ast-1)$ with $\tau=187/91$, and observed that the zero $p_0$ of $\chi$ lies beyond $\pc$ as seen from $1/2$. Our implementation reproduces their tabulated values of $p_0$ and $p^\ast$ for the Archimedean and two-uniform lattices~\cite{Neher08supp} to within $10^{-4}$. On the hold-out set, $p^\ast$ predicts the site thresholds with an rms error of $0.005$ and the bond thresholds with $0.016$; a constant offset fitted on $k\le4$ leaves the site error unchanged and reduces the bond error to $0.011$. The ordering of $p_0$ and $\pc$ holds for 99.8\% of the site thresholds but only for 73\% of the bond thresholds.

\begin{table*}
\caption{Coefficients $\beta_t$ of the linear rule, Eq.~\eqref{eq:mix}: fit on the 243 tilings with $k\le4$ (used for all hold-out results) and, for reference, fit on all 23\,396 tilings. Standard errors, in parentheses, are from a bootstrap over composition groups, since the residuals are correlated within a composition. $^\ast$The type occurs in a single tiling, its Archimedean lattice, so the coefficient equals that threshold and has no meaningful error. Also listed: the number of tilings that contain the vertex type, with $k\le4$ and in total, and the Archimedean thresholds $\pc(t)$ where they exist.}
\label{tab:beta}
\begin{ruledtabular}
\begin{tabular}{lrrllllll}
 & \multicolumn{2}{c}{tilings} & \multicolumn{3}{c}{site} & \multicolumn{3}{c}{bond}\\
type & $k\le4$ & all & $\beta$ ($k\le4$) & $\beta$ (all) & $\ps(t)$ & $\beta$ ($k\le4$) & $\beta$ (all) & $\pb(t)$\\
\hline
$(3^6)$ & 100 & 14286 & 0.5026(7) & 0.5041(2) & 0.5000 & 0.3460(5) & 0.3443(2) & 0.3473 \\
$(3^4.6)$ & 56 & 9903 & 0.5818(6) & 0.5827(1) & 0.5795 & 0.4334(8) & 0.4336(1) & 0.4343 \\
$(3^3.4^2)$ & 86 & 13338 & 0.5505(4) & 0.5508(1) & 0.5502 & 0.4190(4) & 0.4187(1) & 0.4196 \\
$(3^2.4.3.4)$ & 43 & 2086 & 0.5519(8) & 0.5525(2) & 0.5508 & 0.413(1) & 0.4134(2) & 0.4141 \\
$(3^2.4.12)$ & 16 & 666 & 0.716(4) & 0.721(1) &  & 0.596(6) & 0.598(2) &  \\
$(3^2.6^2)$ & 62 & 8476 & 0.656(1) & 0.6539(2) &  & 0.546(1) & 0.5453(2) &  \\
$(3.4.3.12)$ & 12 & 367 & 0.722(10) & 0.723(2) &  & 0.596(14) & 0.590(3) &  \\
$(3.4^2.6)$ & 64 & 12524 & 0.6235(6) & 0.6232(1) &  & 0.5181(6) & 0.5164(2) &  \\
$(3.4.6.4)$ & 36 & 1548 & 0.624(1) & 0.6249(3) & 0.6218 & 0.524(2) & 0.5235(3) & 0.5248 \\
$(3.6.3.6)$ & 76 & 14022 & 0.6502(9) & 0.6521(1) & 0.6527 & 0.5228(9) & 0.5230(2) & 0.5244 \\
$(3.12^2)$ & 12 & 335 & 0.819(15) & 0.831(3) & 0.8079 & 0.759(21) & 0.772(3) & 0.7404 \\
$(4^4)$ & 62 & 12636 & 0.5929(3) & 0.5929(1) & 0.5927 & 0.4980(6) & 0.4982(2) & 0.5000 \\
$(4.6.12)$ & 10 & 381 & 0.757(4) & 0.764(2) & 0.7478 & 0.703(5) & 0.712(1) & 0.6937 \\
$(4.8^2)$ & 1 & 1 & 0.7297$^\ast$ & 0.7297$^\ast$ & 0.7297 & 0.6768$^\ast$ & 0.6768$^\ast$ & 0.6768 \\
$(6^3)$ & 29 & 2146 & 0.701(2) & 0.7058(4) & 0.6970 & 0.655(2) & 0.6560(6) & 0.6527 \\

\end{tabular}
\end{ruledtabular}
\end{table*}

\textit{Composition rules.} For the non-Archimedean tilings whose vertex types all occur in Archimedean lattices, a parameter-free rule is the composition-weighted average of the Archimedean thresholds,
\begin{equation}
\pc\approx\sum_t x_t\,\pc(t),
\label{eq:avg}
\end{equation}
where $x_t$ is the fraction of vertices of type $t$ and $\pc(t)$ is the threshold of the Archimedean lattice of that type. On the hold-out set ``A'' its rms error is $1.5\times10^{-3}$ (site) and $1.7\times10^{-3}$ (bond). To cover all 15 vertex types, including the four without an Archimedean lattice, $(3^2.4.12)$, $(3^2.6^2)$, $(3.4.3.12)$, and $(3.4^2.6)$, we fitted the linear rule
\begin{equation}
\pc\approx\sum_t x_t\,\beta_t
\label{eq:mix}
\end{equation}
with one coefficient per vertex type. The vertex fractions are the only input: the mean degree and the numbers of polygons of each size per vertex are linear combinations of them, so adding these quantities would not change a linear rule. All 15 types occur in the 243 tilings with $k\le4$, although $(4.8^2)$ only in its Archimedean lattice and the types with 12-gons in 10--16 tilings. Fitted on these 243 tilings, the rule has a leave-one-out cross-validation error of $3.6\times10^{-3}$ (site) and $4.9\times10^{-3}$ (bond) over the 242 tilings other than the $(4.8^2)$ lattice, which has leverage one (its coefficient equals its threshold). It predicts the 23\,153 thresholds with $k\ge5$ with an rms error of $1.9\times10^{-3}$ (site) and $2.3\times10^{-3}$ (bond), and the subset ``A'' with $1.0\times10^{-3}$, better than Eq.~\eqref{eq:avg}. The hold-out error decreases with $k$, from $2.7\times10^{-3}$ at $k=5$ to $1.8\times10^{-3}$ at $k=10$ (site) and from $3.6\times10^{-3}$ to $2.2\times10^{-3}$ (bond). Since the hold-out set is dominated by a few large families of mixtures, we also weighted every composition group equally; the hold-out error is then $2.9\times10^{-3}$ (site) and $3.4\times10^{-3}$ (bond). Two variants bracket the linear rule (Table~\ref{tab:rules}). A rule linear in the numbers $f_n$ of $n$-gons per vertex, with five independent parameters (since $\zb=2+2\sum_nf_n$ and $\sum_n(n-2)f_n=2$), has hold-out errors of $2.8\times10^{-3}$ and $4.6\times10^{-3}$, between the Euler estimate and the linear rule. The same linear rule for $\ln[\pc/(1-\pc)]$ reduces the hold-out errors to $1.4\times10^{-3}$ and $2.1\times10^{-3}$, since it describes part of the curvature toward the open end members such as $(3.12^2)$. The rule is thus 2.5 (site) to 7 (bond) times more accurate than the Euler estimate $p^\ast$ and 6--17 times more accurate than the Galam--Mauger formula, at the price of 15 fitted coefficients against none or two. Table~\ref{tab:beta} lists the coefficients of the fit on $k\le4$, used for all hold-out results, and, for reference, of a fit to all 23\,396 tilings. For the common types the coefficients of the full fit are within about $0.004$ of the Archimedean thresholds; the largest differences, up to $0.03$, occur for the types with 12-gons and for $(6^3)$. For $(3.12^2)$ the coefficient exceeds the threshold of the Archimedean lattice of that type by 0.023 (site) and 0.031 (bond), so the linear rule does not reproduce this end member; the coefficients describe the mixtures that dominate the data set, not the pure lattices.

\subsection{Composition and arrangement}

The composition rules leave an rms error of about $2\times10^{-3}$, 30 to 100 times the statistical uncertainty of the atlas. Part of this error is due to the arrangement of the vertices, which a composition rule cannot see. The 23\,396 tilings have 6789 distinct compositions; 3287 of them are shared by two or more tilings (19\,894 tilings in total). For many groups the spread of the thresholds is within the statistical errors (median spread 1.9 combined standard errors for site and 3.2 for bond percolation). A $\chi^2$ test of homogeneity in each group, with the false discovery rate controlled at 1\%~\cite{BenjaminiHochberg95}, finds significant differences in 24\% (site) and 49\% (bond) of the groups. The largest spreads are $5.9\times10^{-3}$ (site) and $2.3\times10^{-2}$ (bond). The largest spreads, in particular for bond percolation, occur for mixtures of $(6^3)$, $(3^2.6^2)$, $(3^4.6)$, and $(3^6)$ vertices, where the honeycomb and triangular regions can be arranged as compact domains or as thin interleaved stripes. Thus composition fixes $\pc$ only to about $10^{-3}$, and the arrangement can shift it by up to almost three orders of magnitude more than the uncertainty of the atlas values. The groups are listed in the Supplemental Material~\cite{SM}.

\textit{Local descriptors.} To see how far the arrangement effect is local, we refine the vertex types into colors of darts (directed edges). The color of a dart leaving vertex $u$ at radius 0 is the sequence of polygons around $u$, starting at the dart; at radius $r$ it adds, in cyclic order, the colors at radius $r-1$ of the darts that point back to $u$ from its neighbors. It is thus determined by the polygons at the vertices within graph distance $r$ of $u$. As descriptors of a tiling we use the fractions of darts of each color, summed over the tiling and its mirror image; at $r=0$ they are equivalent to the composition. Linear rules in these fractions (ridge regression; the ridge parameter is chosen by fitting on $k\le6$ and validating on $k=7$, the rule is then refitted on the 2720 tilings with $k\le7$ and tested on the 20\,676 tilings with $k=8$--10; the numbers of dart colors are 60, 6749, 125\,852, and 289\,765 for $r=0$--3) reduce the rms error from $1.7\times10^{-3}$ at $r=0$ to $0.90\times10^{-3}$ at $r=1$ and $0.75\times10^{-3}$ at $r=2$ for site percolation, and from $2.2\times10^{-3}$ to $1.4\times10^{-3}$ and $1.1\times10^{-3}$ for bond percolation; $r=3$, where most colors occur in few tilings, gives no further gain. Much of this gain is a better description of the dependence on composition. To isolate the arrangement, we centered the thresholds and the descriptors within each of the 3287 composition groups with two or more tilings and fitted only the differences, with five-fold cross-validation over the groups. Descriptors of radius one predict 62\% (site) and 55\% (bond) of the variance of the thresholds within a composition, which has an rms of $2.2\times10^{-4}$ and $6.2\times10^{-4}$; radius two adds little (66\% and 54\%). Statistical errors account for 8\% (site) and 1\% (bond) of this variance, so these fractions are not limited by noise. The arrangement effect is thus to a large part local on the scale of nearest neighbors. This holds for the domains that occur at $k\le10$, which are small and have dense boundaries. For large compact domains of one phase embedded in another, $\pc$ must approach the threshold of the embedding phase, a nonlocal effect that a local rule cannot capture; the largest spreads above, between compact domains and thin stripes, may be a first sign of it.

\section{Slip pairs}\label{sec:slip}

The smallest change of arrangement is a slip: two tilings that share the unit cell content and the first translation $\mathbf T_1$, while $\mathbf T_2'=\mathbf T_2\pm\mathbf T_1/2$. Such tilings consist of the same strips, stacked with different registry; they are the two-dimensional analog of polytypes, crystals built from identical layers in different stacking sequences, and a single slip is a stacking fault~\cite{VermaKrishna66}. Among the tilings with $k\le6$ there are 299 slip pairs (we restricted the census to $k\le6$; the list is in the Supplemental Material~\cite{SM}). Their site thresholds differ by more than five standard errors in 11\% of the pairs and their bond thresholds in 41\%. In all 34 pairs where both differences are resolved at this level, they have the same sign; the family below shows, however, that the site and bond differences can have opposite signs when the strips are far apart. Since many slip differences are unresolved, one may ask whether $\pc$ is exactly invariant under a slip for some pairs, for instance through an isomorphism of the matching graphs.

\begin{figure*}
\includegraphics[width=\textwidth]{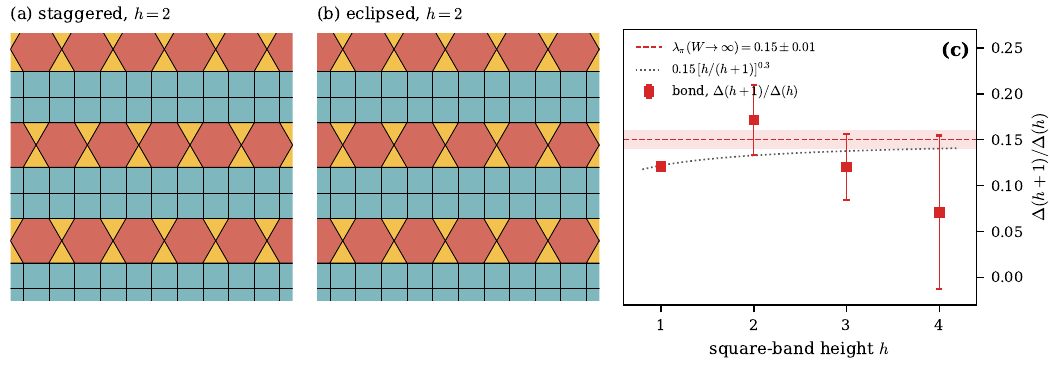}
\caption{(a), (b) The staggered and eclipsed tilings of the slip family with $h=2$ rows of squares between the strips of hexagons and triangles; both have the composition $\tfrac47(3.4^2.6)+\tfrac27(4^4)+\tfrac17(3.6.3.6)$. (c) Ratio of successive bond threshold differences, $\Delta(h+1)/\Delta(h)$ with $\Delta=\pc(\text{eclipsed})-\pc(\text{staggered})$ (Table~\ref{tab:slip}). Dashed line and band: the leading $\kappa=\pi$ eigenvalue of the square-band transfer matrix extrapolated to infinite circumference, $\lambda_\pi=0.15\pm0.01$ (Appendix~\ref{app:tm}). Dotted: ratio for $\Delta\propto h^{-\alpha}\lambda_\pi^h$ with $\alpha=0.3$.}
\label{fig:slip}
\end{figure*}

\subsection{Is a threshold ever invariant under a slip?}

We found no evidence for exact invariance. The two-uniform pair already shows that a slip changes the site threshold: the difference is $2.73(5)\times10^{-4}$, i.e., $51\sigma$. Consistently, we found no isomorphism of the matching graphs that maps the translation lattice of one tiling onto that of the other: for every index from 4 to 8, the sets of quotients of the two matching graphs by all translation sublattices of that index differ (compared with \texttt{nauty}~\cite{McKayPiperno14}). Exact enumeration on small tori is not a useful test. On tori one period wide along $\mathbf T_1$ the two members of a slip pair give identical site polynomials $P_+(p)-P_0(p)$, and for another pair of four-uniform tilings with the same composition the polynomials coincide on all tori we enumerated, although the site and bond thresholds differ by $13\sigma$ and $11\sigma$. Whether some slip pair has exactly equal thresholds therefore remains open; in the family below the differences decay exponentially with the distance between the strips, so that unresolved differences are expected even without an exact symmetry.

The local descriptors of Sec.~\ref{sec:results} show the same trend for all pairs. The two members of a slip pair have equal descriptors up to some radius and different ones from a radius $r^\ast$ on; among the 299 pairs, $r^\ast=1$ for 191, 2 for 67, 3 for 31, 4 for 9, and 5 for one. The bond difference exceeds five standard errors for 120 of the 191 pairs with $r^\ast=1$ (median $|\Delta|=1.7\times10^{-4}$), for 3 of the 67 pairs with $r^\ast=2$, and for none of the 41 pairs with $r^\ast\ge3$; the site difference is resolved only for $r^\ast=1$ (34 pairs). The threshold difference thus falls rapidly with the radius up to which two tilings look alike, as expected for a quasi-local functional of the tiling. In the family below, $r^\ast=1$, 1, 2, 2, and 3 for $h=1$--5, i.e., $r^\ast=\lceil h/2\rceil$: the registry of the strips is first seen from the middle of the band.

\textit{Stackings.} The catalog contains 21 tilings with the composition $\tfrac45(3.4^2.6)+\tfrac15(3.6.3.6)$ of the $h=1$ pair ($k=2$--10). All of them are stackings of the same strips separated by single rows of squares, with one to six strips per period, and differ only in the registry at each strip boundary, eclipsed (E) or staggered (S); for instance, the three-uniform one has the stacking sequence ES. Their thresholds are linear in the fraction $f_E$ of eclipsed boundaries: weighted fits $\pc=\pc^S+f_E(\pc^E-\pc^S)$ give $\chi^2=14.4$ (site) and 15.8 (bond) for 19 degrees of freedom, with largest deviations of 1.9 and 2.4 standard errors. Adding a term proportional to the fraction of neighboring boundaries of different type gives coefficients of $-0.012(16)$ and $-0.005(3)$ in units of $\pc^E-\pc^S$. Each strip boundary thus contributes independently, and interactions between boundaries one strip apart, the analog of the interactions that select polytypes~\cite{VermaKrishna66}, are below about 1\% of the difference between the two registries.

\subsection{A clean family}\label{sec:family}

\begin{table}
\caption{Slip family of Fig.~\ref{fig:slip}: threshold difference $\Delta=\pc(\text{eclipsed})-\pc(\text{staggered})$. Unmarked values are differences of the extrapolated core thresholds. $^c$Coupled simulation of the two members on the same torus (Sec.~\ref{sec:family}): $24\times8$ unit cells for $h=4$ and 5; for site percolation at $h=2$ and 3, the weighted mean of two tori with about 2000 and 8700 sites, which agree within 0.8 and 1.5 combined standard errors.}
\label{tab:slip}
\begin{ruledtabular}
\begin{tabular}{rlrr}
$h$ & $k$-uniform & $\Delta_{\rm site}$ ($10^{-6}$) & $\Delta_{\rm bond}$ ($10^{-6}$)\\
\hline
1 & 2 & $+273(5)$ & $+1218(4)$\\
2 & 3 & $-13.2(1.3)^c$ & $+147(4)$\\
3 & 3 & $+0.3(5)^c$ & $+25(6)$\\
4 & 4 & $-0.3(7)^c$ & $+3.0(6)^c$\\
5 & 4 & $-4(6)$ & $+0.21(25)^c$\\
\end{tabular}
\end{ruledtabular}
\end{table}

To see how the effect of a slip depends on geometry, consider the family of Fig.~\ref{fig:slip}: strips of hexagons and triangle pairs (vertex types $(3.6.3.6)$ inside, $(3.4^2.6)$ on the boundary) separated by bands of $h$ rows of squares, with $(4^4)$ vertices inside the band. The strips have period $|\mathbf T_1|=2$ along their length. In the eclipsed member, consecutive strips are related by a translation perpendicular to them; in the staggered member they are shifted by half a period, one edge length. For $h=1$ the two members are the two two-uniform tilings with composition $\tfrac45(3.4^2.6)+\tfrac15(3.6.3.6)$; for $h=2,3$ they are three-uniform and for $h=4,5$ four-uniform. For $h\ge4$ the differences are below the uncertainty of the core thresholds, so we measured them directly by coupled simulation. The two members of a pair share the cell content, so their sites and bonds can be numbered identically; both are simulated on the same torus $\langle L_x\mathbf T_1,L_y\mathbf T_2\rangle$, which for the staggered member, whose lattice contains $2\mathbf T_2$, requires a twisted boundary condition, and corresponding elements are occupied in the same random order. The variance of $\Delta$ is reduced only by a factor of 2--3 compared with independent runs, because the two configurations differ at every strip boundary and critical crossing events are sensitive to such local changes~\cite{BenjaminiKalaiSchramm99}; the finite-size corrections of the two members, however, largely cancel. Table~\ref{tab:slip} gives the differences and Fig.~\ref{fig:slip}(c) the ratios of successive differences. The bond difference decreases by about an order of magnitude per row of squares, with ratios $\Delta(h+1)/\Delta(h)=0.121(3)$, $0.17(4)$, and $0.12(4)$ for $h=1$--4; $\Delta(5)=0.21(25)\times10^{-6}$ is not resolved. The site difference decreases even faster and changes sign: $\Delta(2)/\Delta(1)=-0.048(5)$, and $\Delta(3)$ to $\Delta(5)$ are not resolved.

\subsection{A transfer-matrix hypothesis}

\begin{table*}
\caption{Leading eigenvalues in the $\kappa=\pi$ sector of the transfer matrix of a square band of circumference $W$ (Appendix~\ref{app:tm}); bond at $p=1/2$, site at $p=0.5927$, the limiting thresholds of the slip family. ``States'' is the number of non-crossing connectivity states (for site percolation an upper bound, Appendix~\ref{app:tm}). At $p=1/2$ the leading bond eigenvalue is twofold degenerate. For site percolation the three largest positive and the two most negative eigenvalues are listed.}
\label{tab:tm}
\begin{ruledtabular}
\begin{tabular}{rrlllrlllll}
 & \multicolumn{4}{c}{bond} & \multicolumn{6}{c}{site}\\
$W$ & states & $\lambda_{\pi,1}$ & $\lambda_{\pi,2}$ & $\lambda_{\pi,3}$ & states & $\lambda_{\pi,1}$ & $\lambda_{\pi,2}$ & $\lambda_{\pi,3}$ & $\lambda_{\pi,-1}$ & $\lambda_{\pi,-2}$\\
\hline
4 & 14 & 0.0882 & 0.0882 & 0.0055 & 51 & 0.0583 & & & & \\
6 & 132 & 0.1034 & 0.1034 & 0.0285 & 731 & 0.0732 & & & $-0.0195$ & $-0.0141$\\
8 & 1430 & 0.1145 & 0.1145 & 0.0760 & 12235 & 0.0823 & 0.0551 & 0.0525 & $-0.0370$ & $-0.0353$\\
10 & 16796 & 0.1226 & 0.1226 & 0.0827 & 223191 & 0.0880 & 0.0647 & 0.0584 & $-0.0484$ & $-0.0469$\\
12 & 208012 & 0.1288 & 0.1288 & 0.0895 & & & & & & \\
\end{tabular}
\end{ruledtabular}
\end{table*}

The slip changes only the relative registry of the strips. Consider the transfer matrix $\mathsf T$ that propagates the connectivity state of a row of the square band by one row (Appendix~\ref{app:tm}). It commutes with the shift $\mathsf S$ by one lattice spacing along the row, so its eigenvectors carry a momentum $\kappa$. A strip acts on the band as a boundary condition with period 2 along the row, i.e., with components in the sectors $\kappa=0$ and $\kappa=\pi$ only. Shifting one strip by one lattice spacing leaves the $\kappa=0$ component unchanged and reverses the sign of the $\kappa=\pi$ component. In terms of the transfer matrix of one period of the tiling in the stacking direction, $\mathsf M=\mathsf C\mathsf T^h$ with $\mathsf C$ for a strip, the staggered tiling has period $\mathsf C\mathsf T^h\,\mathsf C\,\mathsf S\mathsf T^h$ up to a similarity transformation, compared with $(\mathsf C\mathsf T^h)^2$ for the eclipsed one. Since $\mathsf S$ acts as $-1$ on the $\kappa=\pi$ sector, the slip reverses the sign of the $\kappa=\pi$ part $\sum_i\lambda_{\pi,i}^h\mathsf P_{\pi,i}$ of $\mathsf T^h$ in every second band. The spectrum of the period transfer matrix, which determines the threshold, therefore changes at leading order linearly in these terms. Thus, to first order in the propagation of the $\kappa=\pi$ component through the band,
\begin{equation}
\Delta(h)\simeq\sum_i A_i\lambda_{\pi,i}^{\,h},
\label{eq:tm}
\end{equation}
where $\lambda_{\pi,i}$ are the eigenvalues of $\mathsf T$ in the $\kappa=\pi$ sector and the amplitudes $A_i$ depend on the strips. For $h=1$ and 2 the coupling between the strips is not weak, so the argument is heuristic there. We evaluate $\mathsf T$ at the limiting thresholds of the family as $h\to\infty$, $p=1/2$ (bond) and 0.5927 (site), where the band is critical, as the hypothesis below requires. At the thresholds of the tilings studied, $p\approx0.51$ and 0.62, the bond eigenvalues differ by less than $5\times10^{-4}$ and the site eigenvalues are about 7\% smaller (Appendix~\ref{app:tm}). Long-wavelength fluctuations live in the sector $\kappa\to0$; in the $\kappa=\pi$ sector, which describes a modulation on the lattice scale, all eigenvalues we computed are small, $|\lambda_\pi|<0.15$, so $\Delta$ is expected to decay exponentially with $h$.

For bond percolation at $p=1/2$, the leading $\kappa=\pi$ eigenvalue on a cylinder of circumference $W$ is $\lambda_\pi=0.088$, 0.103, 0.114, 0.123, and 0.129 for $W=4$ to 12 (Table~\ref{tab:tm}). The increments decrease geometrically (ratio $\approx0.75$), and extrapolations in $1/W$, $1/W^2$, or by a geometric tail give $\lambda_\pi(W\to\infty)=0.15\pm0.01$. At $p=1/2$ the leading eigenvalue is twofold degenerate for every $W$, presumably a consequence of self-duality; at $p=0.51$ the pair splits slightly (0.1292 and 0.1280 at $W=12$). Further $\kappa=\pi$ eigenvalues follow at 0.090 and 0.074. The measured ratios, 0.121(3), 0.17(4), and 0.12(4), are thus of the same magnitude as $\lambda_\pi$. A weighted fit $\Delta(h)=A\lambda^h$ to all five bond values gives $\lambda=0.123(3)$ ($\chi^2=3.6$ for three degrees of freedom), dominated by $h=1$ and 2; this is 2.6 combined standard errors below $\lambda_\pi=0.15\pm0.01$. Without $h=1$, the most strongly coupled case, where a first-order argument is least reliable, the fit gives $\lambda=0.142$ (68\% interval 0.129--0.153, $\chi^2=1.3$ for two degrees of freedom), in agreement with $\lambda_\pi$. The first ratio lies below $\lambda_\pi$. A ratio that increases toward $\lambda_\pi$ with $h$ follows from Eq.~\eqref{eq:tm} if subleading modes contribute with the same sign at small $h$. It also follows if, at infinite circumference, $\lambda_\pi$ is the edge of a continuum of eigenvalues, which the spectra for $W\le12$ can neither confirm nor exclude; then $\Delta(h)\propto h^{-\alpha}\lambda_\pi^h$, the ratios approach $\lambda_\pi$ from below, and the rate $\lim h^{-1}\ln|\Delta|$ is unchanged. With $\lambda_\pi=0.15$ fixed, this form with $\alpha=0.30$ describes all five values ($\chi^2=1.8$ for three degrees of freedom), whereas a pure exponential does not ($\chi^2=89$ for four). Our data cannot distinguish between subleading modes and a continuum edge. For site percolation at $p=0.5927$, the $\kappa=\pi$ sector contains positive and negative eigenvalues of similar magnitude: for $W=10$, $0.088$, $0.065$, $0.058$, $-0.047$, and $-0.048$. A sign change of $\Delta$ is therefore allowed, and the measured ratio $-0.048(5)$ is consistent with a dominant contribution of the negative modes. With only two resolved values, however, the amplitudes cannot be determined independently, so the site data are consistent with Eq.~\eqref{eq:tm} but do not test it.

These observations motivate the following hypothesis, which the present data support but do not establish. Consider tilings composed of fixed strips separated by bands of $h$ rows of a periodic lattice, and two registries of the strips that differ by a shift of momentum $\kappa_0$. Then $\lim_{h\to\infty}h^{-1}\ln|\Delta\pc(h)|=\ln|\lambda|$, where $\lambda$ is the eigenvalue of largest modulus of the infinite-circumference band transfer matrix in the sector $\kappa_0$ that has a nonzero amplitude, evaluated at the limiting threshold of the family. For the square band and $\kappa_0=\pi$ this predicts that the bond difference eventually falls by a factor of about 7 per row, possibly with a power-law prefactor. The hypothesis can be tested on families with larger $h$ and on other strip types, which would also check that the rate does not depend on the strips; larger $h$ requires a stronger variance reduction than our coupled simulation provides. A proof would require control of the critical model on the whole tiling beyond the first-order argument above.

\section{Conclusions}\label{sec:concl}

We have computed the site and bond percolation thresholds of all 23\,396 $k$-uniform tilings with $k\le10$, after verifying exactly that the periodic graphs are pairwise distinct tilings with $k$ vertex orbits. For $k\le4$ the values have uncertainties of a few units in the sixth decimal and are validated against all 22 Archimedean reference values; to our knowledge, they include the first thresholds of the 61 three-uniform and 151 four-uniform tilings and the first published bond thresholds of the two-uniform lattices with error estimates, which agree with unpublished high-precision values~\cite{GuZiff} to $10^{-6}$, and they revise one two-uniform site threshold. The estimator based on the equality of the no-wrapping and cross-wrapping probabilities, the Monte Carlo form of the critical-polynomial criterion, requires no knowledge of the torus anisotropy and is therefore suited to lattices of low symmetry. For the atlas values for $k=5$--10, a direct check on 260 thresholds, including the tilings with the largest cells, the most elongated tori, and the largest square domains, bounds the finite-size bias well below their statistical errors. An exact test on small tori confirms the simulation code without any extrapolation.

On the hold-out tilings, the vertex composition alone fixes the threshold to about $2\times10^{-3}$ through a rule fitted within the family; the published universal formulas~\cite{GalamMauger96,Neher08,Lebrecht21,GalamMauger98}, which are not fitted to this family, reach $5\times10^{-3}$ at best (Table~\ref{tab:rules}). Beyond this level the arrangement matters, by up to $6\times10^{-3}$ for site and $2\times10^{-2}$ for bond percolation. For the small domains that occur at $k\le10$, descriptors of the neighborhood of radius one predict more than half of the variance of the thresholds within a composition, stackings of identical strips have thresholds linear in the fractions of the two boundary types, and slip pairs whose members agree up to larger radii have rapidly smaller threshold differences: in this regime the threshold behaves as a quasi-local functional of the tiling. For the smallest arrangement change, a slip of strips, the bond difference decays exponentially with the distance between the strips, at a rate of the same magnitude as the leading eigenvalue of the momentum-$\pi$ sector of the band transfer matrix; for $h\ge2$ the fitted rate, $0.142$ (68\% interval 0.129--0.153), is consistent with it. We formulate the relation as a hypothesis.

The data invite several extensions. Critical polynomials computed by transfer matrices~\cite{Jacobsen14,ScullardJacobsen20} could improve the core values by several digits and provide high-precision site thresholds of lattices with several vertex orbits, which are still missing as references. Tilings composed of squares and triangles only, which approach dodecagonal quasicrystals~\cite{OxborrowHenley93}, could be followed to large $k$ and compared with the thresholds of quasiperiodic tilings~\cite{Babalievski95,ZiffBabalievski99}. The same lattices could also serve as test cases for the Potts model, where the corresponding wrapping identity contains the factor $q$~\cite{Arguin02,JacobsenScullard13}.

\begin{acknowledgments}
This research received no external funding. The author declares no competing interests. The author chose the research problem, directed the work and the computations, decided on the scope of the results and on their presentation, checked the results, and takes full responsibility for the content of the article. The large language model Claude Opus~5.5 (Anthropic) was employed under the strict control of the author and within a limited scope set by the author. Within this scope it was used to draft and edit the computer code for the computations and the text of the manuscript, to speed up the work. The author specified the computations and the checks and reviewed the code and the text; all results were verified by the independent tests described in the paper, namely exact relations, reference values, repeated and independent simulations, and the exact verification of the periodic graphs.
\end{acknowledgments}

\section*{Data availability}
The thresholds of all 23\,396 tilings, with their uncertainties, compositions, torus data, and the periodic graphs, as well as the lists of slip pairs and composition groups and the results of the atlas finite-size check, are provided in the Supplemental Material~\cite{SM}. The periodic graphs are derived from the catalog snapshot of Ref.~\cite{TesCatalog} (commit 6772c7c6). The source code (MIT License), the job lists with their random seeds, the processed results (CC BY 4.0), and the scripts that generate all tables and figures from them are openly available at \url{https://github.com/Hardos220/percolation-k-uniform-tilings} and are archived, together with the raw histograms of $n_e$ and $n_x$ of all runs, at Zenodo, \href{https://doi.org/10.5281/zenodo.23533708}{doi:10.5281/zenodo.23533708}.

\appendix

\section{Transfer matrix of the square band}\label{app:tm}

The band is a cylinder of the square lattice with circumference $W$ (even, to accommodate the period-2 boundary). The state of a row is a partition of its $W$ sites according to connectivity through the part of the band already traversed; for site percolation the state also records which sites are occupied, and only occupied sites carry connectivity labels. The traversed part of the band is an annulus bounded by the current row. A path within it that joins two sites of the row cuts off, together with one of the two arcs of the row between these sites, a disk, so no cluster can join a site of this arc with a site of the other arc; hence only partitions that are non-crossing in the cyclic order occur. Their numbers are the Catalan numbers $C_W$ for bond percolation. For site percolation the enumerated states include some that cannot occur, such as adjacent occupied sites in different blocks; they contribute only zero eigenvalues, and the numbers of states in Table~\ref{tab:tm} are upper bounds. One step of $\mathsf T$ adds a row of squares: first the vertical links to the new row (bond: each present with probability $p$; site: each new site occupied with probability $p$), then the horizontal links within the new row (bond: each present with probability $p$; site: present between occupied neighbors). For bond percolation this is the connectivity transfer matrix of the $q\to1$ Potts model~\cite{TemperleyLieb71,JacobsenScullard13}. We first computed the spectra in the basis of all set partitions; since $\mathsf T$ maps non-crossing states to non-crossing states, the matrix is block-triangular in that basis, and the spectrum of the non-crossing block (Table~\ref{tab:tm}) coincides, to all digits shown, with the leading eigenvalues found in the full basis for $W\le10$. The matrix is column-stochastic, so its largest eigenvalue in the $\kappa=0$ sector is 1. Since $\mathsf T$ commutes with the shift $\mathsf S$, the projector $\frac1W\sum_j e^{-i\kappa j}\mathsf S^j$ onto momentum $\kappa$ commutes with $\mathsf T$, and the eigenvalues in the $\kappa=\pi$ sector are obtained by the Arnoldi method applied to $\mathsf T$ times this projector. We use $p=1/2$ (bond) and 0.5927 (site), the thresholds of the square lattice, which are the limiting thresholds of the slip family. At $p=0.51$ and 0.62, close to the thresholds of the $h=1$ tilings, the bond eigenvalues differ from those of Table~\ref{tab:tm} by less than $5\times10^{-4}$ and the site eigenvalues are about 7\% smaller.

\section{Three-uniform tilings}\label{app:tab}

Tables~\ref{tab:3uni} and \ref{tab:3unib} list the thresholds of the 61 three-uniform tilings. The identifier is the name of the file in the catalog of Ref.~\cite{TesCatalog}; it encodes the vertex orbits.

\begin{table*}
\caption{Site and bond thresholds of the 61 three-uniform tilings, ordered by decreasing site threshold (continued in Table~\ref{tab:3unib}). ID: catalog file name~\cite{TesCatalog}; $\zb$: mean degree.}
\label{tab:3uni}
\footnotesize
\begin{ruledtabular}
\begin{tabular}{rllrll}
 & ID & vertex types & $\zb$ & $\ps$ & $\pb$\\
\hline
1 & \texttt{3a 4a 4c 1} & $2/5\,(3^{2}.4.12)+2/5\,(3.4.3.12)+1/5\,(3.12^{2})$ & 3.800 & 0.7297711(25) & 0.6190292(27) \\
2 & \texttt{3a 4c 4l 1} & $1/2\,(3.4.3.12)+1/4\,(3.12^{2})+1/4\,(3.4.6.4)$ & 3.750 & 0.7255320(27) & 0.6224234(29) \\
3 & \texttt{3a 4a 4l 1} & $1/2\,(3^{2}.4.12)+1/4\,(3.12^{2})+1/4\,(3.4.6.4)$ & 3.750 & 0.7243694(26) & 0.6300425(30) \\
4 & \texttt{3c 4a 6d 1} & $6/13\,(3^{2}.4.12)+6/13\,(4.6.12)+1/13\,(3^{6})$ & 3.692 & 0.7166544(38) & 0.6336893(33) \\
5 & \texttt{3c 4h 4k 1} & $2/5\,(3.4^{2}.6)+2/5\,(4.6.12)+1/5\,(3.6.3.6)$ & 3.600 & 0.6857748(35) & 0.5911238(28) \\
6 & \texttt{3f 4d 4f 1} & $2/5\,(3^{2}.6^{2})+2/5\,(6^{3})+1/5\,(3.6.3.6)$ & 3.600 & 0.6697939(32) & 0.5821763(27) \\
7 & \texttt{3f 4d 4h 1} & $1/3\,(3^{2}.6^{2})+1/3\,(3.6.3.6)+1/3\,(6^{3})$ & 3.667 & 0.6671122(33) & 0.5713593(27) \\
8 & \texttt{4d 4g 4h 1} & $2/3\,(3.6.3.6)+1/3\,(3^{2}.6^{2})$ & 4.000 & 0.6512046(38) & 0.5300788(34) \\
9 & \texttt{3d 4d 6a 1} & $2/3\,(3^{2}.6^{2})+2/9\,(6^{3})+1/9\,(3^{6})$ & 4.000 & 0.6511325(36) & 0.5544924(30) \\
10 & \texttt{3f 4d 5a 1} & $1/3\,(3^{4}.6)+1/3\,(3^{2}.6^{2})+1/3\,(6^{3})$ & 4.000 & 0.6459708(61) & 0.5417084(45) \\
11 & \texttt{4a 4m 5c 1} & $2/5\,(3^{2}.4.12)+2/5\,(3.4.6.4)+1/5\,(3^{3}.4^{2})$ & 4.200 & 0.6437793(39) & 0.5305853(32) \\
12 & \texttt{4g 4h 4k 2} & $1/2\,(3.4^{2}.6)+1/2\,(3.6.3.6)$ & 4.000 & 0.6376763(50) & 0.5210539(43) \\
13 & \texttt{4g 4h 4k 1} & $1/2\,(3.4^{2}.6)+1/2\,(3.6.3.6)$ & 4.000 & 0.6375194(53) & 0.5202821(43) \\
14 & \texttt{4a 5e 6f 1} & $4/7\,(3^{2}.4.12)+2/7\,(3^{2}.4.3.4)+1/7\,(3^{6})$ & 4.571 & 0.6373104(53) & 0.4985799(34) \\
15 & \texttt{4d 4h 4k 2} & $1/2\,(3.4^{2}.6)+1/4\,(3^{2}.6^{2})+1/4\,(3.6.3.6)$ & 4.000 & 0.6363470(25) & 0.5253072(24) \\
16 & \texttt{4d 4h 4k 1} & $1/2\,(3.4^{2}.6)+1/4\,(3^{2}.6^{2})+1/4\,(3.6.3.6)$ & 4.000 & 0.6361815(64) & 0.5244735(27) \\
17 & \texttt{4h 4k2 6} & $4/5\,(3.4^{2}.6)+1/5\,(3.6.3.6)$ & 4.000 & 0.6282443(34) & 0.5182987(29) \\
18 & \texttt{4a 5c 6i 1} & $1/2\,(3^{2}.4.12)+1/4\,(3^{6})+1/4\,(3^{3}.4^{2})$ & 4.750 & 0.6240755(55) & 0.4872769(34) \\
19 & \texttt{4k 4l2 1} & $1/2\,(3.4^{2}.6)+1/2\,(3.4.6.4)$ & 4.000 & 0.6218128(35) & 0.5205844(30) \\
20 & \texttt{4e 4k 5c 1} & $2/5\,(3^{2}.6^{2})+2/5\,(3.4^{2}.6)+1/5\,(3^{3}.4^{2})$ & 4.200 & 0.6210445(52) & 0.5094462(44) \\
21 & \texttt{4f 4k 4r 1} & $4/7\,(3.4^{2}.6)+2/7\,(4^{4})+1/7\,(3.6.3.6)$ & 4.000 & 0.6183577(32) & 0.5130626(27) \\
22 & \texttt{4f 4k 4s 1} & $4/7\,(3.4^{2}.6)+2/7\,(4^{4})+1/7\,(3.6.3.6)$ & 4.000 & 0.6183478(32) & 0.5132098(27) \\
23 & \texttt{4k 4l 4o 1} & $4/7\,(3.4^{2}.6)+2/7\,(3.4.6.4)+1/7\,(4^{4})$ & 4.000 & 0.6176267(18) & 0.5159154(16) \\
24 & \texttt{4f 4k 4u 1} & $4/9\,(3.4^{2}.6)+4/9\,(4^{4})+1/9\,(3.6.3.6)$ & 4.000 & 0.6128031(37) & 0.5102344(32) \\
25 & \texttt{4f 4k 4u 2} & $4/9\,(3.4^{2}.6)+4/9\,(4^{4})+1/9\,(3.6.3.6)$ & 4.000 & 0.6127977(61) & 0.5102596(46) \\
26 & \texttt{4h 5a 6a 1} & $6/13\,(3^{4}.6)+6/13\,(3.6.3.6)+1/13\,(3^{6})$ & 4.615 & 0.6105655(18) & 0.4729273(17) \\
27 & \texttt{4k 4l 5e 1} & $1/2\,(3.4^{2}.6)+1/4\,(3^{2}.4.3.4)+1/4\,(3.4.6.4)$ & 4.250 & 0.6055637(17) & 0.4914115(16) \\
28 & \texttt{4d 5a 6f 1} & $2/5\,(3^{4}.6)+2/5\,(3^{2}.6^{2})+1/5\,(3^{6})$ & 4.800 & 0.5969555(24) & 0.4605932(35) \\
29 & \texttt{4e 5b 6i 3} & $2/5\,(3^{4}.6)+2/5\,(3^{2}.6^{2})+1/5\,(3^{6})$ & 4.800 & 0.5962765(57) & 0.4592692(45) \\
30 & \texttt{4f 5a 6a 1} & $3/5\,(3^{4}.6)+3/10\,(3.6.3.6)+1/10\,(3^{6})$ & 4.800 & 0.5957901(49) & 0.4548427(30) \\
31 & \texttt{4k 5e 6a 1} & $12/19\,(3.4^{2}.6)+6/19\,(3^{2}.4.3.4)+1/19\,(3^{6})$ & 4.421 & 0.5957781(38) & 0.4761811(28) \\

\end{tabular}
\end{ruledtabular}
\end{table*}

\begin{table*}
\caption{Continuation of Table~\ref{tab:3uni}.}
\label{tab:3unib}
\footnotesize
\begin{ruledtabular}
\begin{tabular}{rllrll}
 & ID & vertex types & $\zb$ & $\ps$ & $\pb$\\
\hline
32 & \texttt{4g 5b2 1} & $4/5\,(3^{4}.6)+1/5\,(3.6.3.6)$ & 4.800 & 0.5951624(36) & 0.4532446(29) \\
33 & \texttt{4m 5e 6a 1} & $12/19\,(3.4.6.4)+6/19\,(3^{2}.4.3.4)+1/19\,(3^{6})$ & 4.421 & 0.5951617(40) & 0.4785771(27) \\
34 & \texttt{4i 5b2 1} & $4/5\,(3^{4}.6)+1/5\,(3.6.3.6)$ & 4.800 & 0.5951600(54) & 0.4532491(41) \\
35 & \texttt{4k 5a 5c 1} & $1/2\,(3.4^{2}.6)+1/4\,(3^{4}.6)+1/4\,(3^{3}.4^{2})$ & 4.500 & 0.5949609(17) & 0.4734438(27) \\
36 & \texttt{4d 5a 6h 1} & $1/3\,(3^{6})+1/3\,(3^{4}.6)+1/3\,(3^{2}.6^{2})$ & 5.000 & 0.5826779(35) & 0.4420118(29) \\
37 & \texttt{4s2 5c 1} & $2/3\,(4^{4})+1/3\,(3^{3}.4^{2})$ & 4.333 & 0.5788109(29) & 0.4718739(21) \\
38 & \texttt{4p 4s 5c 1} & $3/5\,(4^{4})+2/5\,(3^{3}.4^{2})$ & 4.400 & 0.5759992(32) & 0.4664677(26) \\
39 & \texttt{4f 5b 6i 1} & $4/7\,(3^{4}.6)+2/7\,(3^{6})+1/7\,(3.6.3.6)$ & 5.143 & 0.5700176(69) & 0.4227173(26) \\
40 & \texttt{4l 5d 6d 1} & $3/5\,(3^{3}.4^{2})+3/10\,(3.4.6.4)+1/10\,(3^{6})$ & 4.800 & 0.5690721(39) & 0.4437350(27) \\
41 & \texttt{4l 5f 6d 1} & $3/5\,(3^{2}.4.3.4)+3/10\,(3.4.6.4)+1/10\,(3^{6})$ & 4.800 & 0.5689262(38) & 0.4371290(26) \\
42 & \texttt{5a 5b 6f 1} & $6/7\,(3^{4}.6)+1/7\,(3^{6})$ & 5.143 & 0.5687854(26) & 0.4213160(27) \\
43 & \texttt{4s 5c2 2} & $2/3\,(3^{3}.4^{2})+1/3\,(4^{4})$ & 4.667 & 0.5646654(26) & 0.4453207(22) \\
44 & \texttt{5a 5c 5e 1} & $1/3\,(3^{4}.6)+1/3\,(3^{3}.4^{2})+1/3\,(3^{2}.4.3.4)$ & 5.000 & 0.5601703(25) & 0.4224297(15) \\
45 & \texttt{4p 5c2 1} & $4/5\,(3^{3}.4^{2})+1/5\,(4^{4})$ & 4.800 & 0.5589115(32) & 0.4349439(24) \\
46 & \texttt{4s 5c 6f 1} & $2/5\,(3^{3}.4^{2})+2/5\,(4^{4})+1/5\,(3^{6})$ & 4.800 & 0.5583123(42) & 0.4348991(28) \\
47 & \texttt{4o 5d 5f 1} & $4/9\,(3^{3}.4^{2})+4/9\,(3^{2}.4.3.4)+1/9\,(4^{4})$ & 4.889 & 0.5552546(37) & 0.4248748(25) \\
48 & \texttt{5d 5f2 3} & $2/3\,(3^{2}.4.3.4)+1/3\,(3^{3}.4^{2})$ & 5.000 & 0.5505561(40) & 0.4155871(27) \\
49 & \texttt{5d2 5f 1} & $2/3\,(3^{3}.4^{2})+1/3\,(3^{2}.4.3.4)$ & 5.000 & 0.5503553(42) & 0.4173329(27) \\
50 & \texttt{4p 5c 6f 1} & $1/2\,(3^{3}.4^{2})+1/4\,(3^{6})+1/4\,(4^{4})$ & 5.000 & 0.5494188(38) & 0.4197311(26) \\
51 & \texttt{4s 5c 6i 1} & $1/3\,(3^{6})+1/3\,(3^{3}.4^{2})+1/3\,(4^{4})$ & 5.000 & 0.5491069(48) & 0.4197692(32) \\
52 & \texttt{5e2 6a 1} & $12/13\,(3^{2}.4.3.4)+1/13\,(3^{6})$ & 5.077 & 0.5471092(19) & 0.4085816(15) \\
53 & \texttt{5c 5e 6a 1} & $6/13\,(3^{3}.4^{2})+6/13\,(3^{2}.4.3.4)+1/13\,(3^{6})$ & 5.077 & 0.5466653(17) & 0.4102233(15) \\
54 & \texttt{5a 6c 6f 1} & $6/11\,(3^{4}.6)+5/11\,(3^{6})$ & 5.455 & 0.5444742(35) & 0.3931541(27) \\
55 & \texttt{5c2 6f 1} & $4/5\,(3^{3}.4^{2})+1/5\,(3^{6})$ & 5.200 & 0.5407132(35) & 0.4047014(26) \\
56 & \texttt{4p 5c 6i 1} & $2/5\,(3^{6})+2/5\,(3^{3}.4^{2})+1/5\,(4^{4})$ & 5.200 & 0.5399914(40) & 0.4048875(27) \\
57 & \texttt{5c2 6i 2} & $2/3\,(3^{3}.4^{2})+1/3\,(3^{6})$ & 5.333 & 0.5342092(33) & 0.3949105(27) \\
58 & \texttt{5a 6f 6i 1} & $3/5\,(3^{6})+2/5\,(3^{4}.6)$ & 5.600 & 0.5328439(34) & 0.3804784(26) \\
59 & \texttt{5b 6j2 1} & $2/3\,(3^{6})+1/3\,(3^{4}.6)$ & 5.667 & 0.5274497(34) & 0.3748076(25) \\
60 & \texttt{5c 6f 6i 1} & $3/5\,(3^{6})+2/5\,(3^{3}.4^{2})$ & 5.600 & 0.5209515(33) & 0.3756874(26) \\
61 & \texttt{5c 6i2 1} & $2/3\,(3^{6})+1/3\,(3^{3}.4^{2})$ & 5.667 & 0.5175410(38) & 0.3709436(26) \\

\end{tabular}
\end{ruledtabular}
\end{table*}

\clearpage
\bibliography{refs}

\end{document}